\documentclass[useAMS,usenatbib]{mn2e}

\usepackage{amssymb, amsmath, amsfonts}

\usepackage{graphicx, shadow}

\usepackage{epstopdf}

\usepackage{rotating}
\usepackage{wasysym}
\usepackage{verbatim}
\usepackage{natbib}

\newcommand{\sunrise}{\textsc{Sunrise}}

\newcommand{\art}{{ART}}
\newcommand{\yt}{{\small yt}}

\newcommand{\hst}{{\sl HST}}

\newcommand{\candels}{{CANDELS}}

\newcommand{\se}{\textsc{SExtractor}}
\newcommand{\gmtwenty}{$G$-$M_{20}$}
\newcommand{\ginimtwenty}{Gini-$M_{20}$}
\newcommand{\sersic}{S\'{e}rsic}

\newcommand{\mum}{\ensuremath{\mu m}}

\title[Structural Evolution in Mock Images]{Diverse Structural Evolution at $z > 1$ in Cosmologically Simulated Galaxies}

\author[G.F.\ Snyder et al.]{Gregory F. Snyder$^1$, Jennifer Lotz$^1$, Christopher Moody$^{2}$, Michael Peth$^3$, \newauthor
Peter Freeman$^4$, Daniel Ceverino$^5$, Joel Primack$^2$, Avishai Dekel$^6$ \\
$^1$ Space Telescope Science Institute, 3700 San Martin Dr, Baltimore, MD 21218 \\
$^2$ Department of Physics, University of California, Santa Cruz, 1156 High Street, Santa Cruz, CA 95064 \\
$^3$ Department of Physics and Astronomy, Johns Hopkins University, 3400 N. Charles Street, Baltimore, MD 21218 \\
$^4$ Department of Statistics, Carnegie Mellon University, 5000 Forbes Avenue, Pittsburgh, PA 15213, USA \\
$^5$ Departamento de F\'{i}sica Te\'{o}rica, Universidad Aut\'{o}noma de Madrid, 28049 Madrid, Spain \\
$^6$ Racah Institute of Physics, The Hebrew University, Jerusalem 91904 Israel}
\begin{document}

\label{firstpage}

\maketitle

\begin{abstract}

From mock Hubble Space Telescope images, we quantify non-parametric statistics of galaxy morphology, thereby predicting the emergence of relationships among stellar mass, star formation, and observed rest-frame optical structure at $1 < z < 3$.  We measure automated diagnostics of galaxy morphology in cosmological simulations of the formation of 22 central galaxies with $9.3 < \log_{10} M_*/M_{\odot} < 10.7$.  These high-spatial-resolution zoom-in calculations enable accurate modeling of the rest-frame UV and optical morphology. Even with small numbers of galaxies, we find that structural evolution is neither universal nor monotonic: galaxy interactions can trigger either bulge or disc formation, and optically bulge-dominated galaxies at this mass may not remain so forever. Simulated galaxies with $M_* > 10^{10} M_{\odot}$ contain relatively more disc-dominated light profiles than those with lower mass, reflecting significant disc brightening in some haloes at $1 < z < 2$.  By this epoch, simulated galaxies with specific star formation rates below $10^{-9.7}\rm\ yr^{-1}$ are more likely than normal star-formers to have a broader mix of structural types, especially at $M_* > 10^{10} M_{\odot}$.  We analyze a cosmological major merger at $z\sim 1.5$ and find that the newly proposed MID morphology diagnostics trace later merger stages while \ginimtwenty\ trace earlier ones.  MID is sensitive also to clumpy star-forming discs.  The observability time of typical MID-enhanced events in our simulation sample is $< 100$ Myr.  A larger sample of cosmological assembly histories may be required to calibrate such diagnostics in the face of their sensitivity to viewing angle, segmentation algorithm, and various phenomena such as clumpy star formation and minor mergers.  

\end{abstract}


\section{Introduction} \label{s:intro}

A galaxy's morphology reflects structures that formed through complex and dynamic assembly processes.   We obtain clues to these galaxy formation physics by measuring this morphology in large numbers of galaxies over cosmic time.   Such surveys have demonstrated a tight connection between structural evolution and its implications for galaxy formation through other properties such as galaxy mass and star formation rate \citep[e.g.,][]{Kauffmann2003,Wuyts2011,Omand2014}.  However, since we cannot directly watch galaxies evolve, the causal relationship between their formation physics and observed morphology is impossible to determine directly.  

Hydrodynamical simulations of evolving galaxies allow us to calibrate these diagnostics by measuring their observability given a set of formation scenarios and physical processes \citep[e.g.,][]{Jonsson:2006, Rocha2007, lotz08, bush10,narayanan10_smg, Hayward2012,Snyder2013,Lanz2014}.  The quality and breadth of these experiments are limited by the availability of computational resources and the fidelity of models for galaxy physics such as star formation, supernovae, and the interstellar medium (ISM).  It has only recently become widespread to model the formation of galaxies \emph{ab initio} \citep[e.g.,][]{Governato2004,Agertz2011,Guedes2011,Marinacci2013,Ceverino2014}, and the realism continues to improve \citep{Stinson2012,Hopkins2013a,Torrey2014}, albeit with still widely varying physics models \citep[e.g.,][]{Scannapieco2012,Kim2014}.  Prior to these advances, studies were limited to small numbers of isolated galaxies or mergers to inform common diagnostics of galaxy evolution, an approach with a significant limitation: they do not fully account for cosmological context, such as gas accretion and the breadth of assembly histories.  In addition to mergers, models of high redshift galaxy formation \citep[e.g.,][]{Dekel2009,Dekel2013} have recently appreciated the tight coupling between gas accretion and disc evolution \citep[e.g.,][]{Danovich2012,Cacciato2012,Dekel2013b}, as well as bulge and SMBH growth mediated by turbulent motions or violent disc instability \citep[e.g.,][]{Bournaud2011, Porter2014} and the evolution of giant clumps \citep{Dekel2013a}.  These important processes likely complicate interpretation of a given observation, and recent studies of galaxy morphology have begun to exploit simulations including them \citep[e.g.,][]{scannapieco10, Pedrosa2014}.  

With the advent of deep, near-IR surveys and improving models, it is now possible to test the physics responsible for the emergence and evolution of galaxy structures in the early universe. In this paper, we analyze rest-frame optical morphologies of galaxies forming in very high spatial resolution cosmological simulations, as if they were observed by the Hubble Space Telescope (HST), quantifying their dependence on time, mass, and star formation rate.   Section~\ref{s:simages} summarizes the hydrodynamical and dust radiative transfer simulations, their conversion into realistic mock HST images, and the pipeline for measuring photometry and non-parametric morphology, using analysis routines identical to those applied to distant galaxy surveys with \hst\ such as \candels.  Section~\ref{s:morphs} presents the predicted morphological evolution at $1 < z < 4$, using non-parametric quantitative measures, for simulated galaxies with $M_* \sim 10^{10} M_{\odot}$.  We discuss implications of this evolution in Section~\ref{s:discussion} and conclude in Section~\ref{s:conclusion}.


\section{Simulations and Mock Images} \label{s:simages}

\citet{Moody2014} presented basic details of the mock data pipeline, and a series of papers presented the full details of the zoom-in simulations \citep{Ceverino2009,Ceverino2010,Ceverino2012,Dekel2013,Ceverino2014} .  We summarize here the elements essential to our morphological analysis.  

\subsection{Hydrodynamical Simulations} \label{ss:hydroart}

We analyze $32$ cosmological hydrodynamical simulations calculated with the Eulerian gasdynamcs + N-body Adaptive Refinement Tree code \citep[\art,][]{Kravtsov1997,Kravtsov2003} in a WMAP5 cosmology \citep{Komatsu2009}.  In addition to gravity and gasdynamics, we apply sub-grid models for various physical processes, following \citet{Ceverino2009}.  They include treatments for the physics of gas and metal cooling, UV-background photoionization, stochastic star formation, gas recycling and metal enrichment, and thermal feedback from supernovae \citep{Ceverino2010,Ceverino2012}, plus a new implementation of feedback from radiation pressure and radiative heating by young stars \citep[][hereafter C14]{Ceverino2014}.  These have been used to understand the gasdynamical processes affecting the evolution of galaxy sizes and hence the emergence of compact red and blue galaxies at high redshift \citep{Zolotov2014}.  The simulations we analyze here are organized into 10 pairs, where each pair consists of a simulation with and without this new source of feedback (RP and no-RP), plus 12 simulations for which only the no-RP simulations were analyzed.   At the masses we consider, the RP feedback models are roughly $50\%$ closer than the non-RP simulations to the observed stellar mass-halo mass relation, achieving agreement to within roughly a factor of two in halo mass \citep[e.g.,][]{Behroozi2013}.  

Collisionless star particles form out of the low-temperature gas in a Monte Carlo fashion, following a prescription that recovers the observed relationship between gas surface density and star formation rate \citep{Kennicutt:1998}.  For our purpose, the essential outputs of the simulation are the three-dimensional positions, masses, and ages of these stochastically spawned star particles, and the three-dimensional metal mass density.  The finest adaptive mesh refinement scale of the latter is $17$--$35\rm\ pc$, while the greatest refinement scale is $108\rm\ pc$ comoving.  Star particles are typically $M \sim 10^5 M_{\odot}$, and dark matter particles are $8\times 10^4 M_{\odot}$.  See \citet{Ceverino2009} and \citet{Ceverino2014} for full details.

We initialized the zoom-in hydrodynamical simulations by randomly selecting haloes with virial masses $10^{11} < M/M_{\odot} < 10^{12}$ and no ongoing major merger at $z = 0.8$ from a coarsely resolved N-body simulation $\sim 30\rm\ Mpc$ across.  We then re-sampled and re-simulated those regions in full hydrodynamics at high resolution with the physics models described above, in all cases to $z \sim 1$ and in some cases to $z \sim 0.7$.    The galaxies forming in these simulations have $9.3 < \log_{10} M_*/M_{\odot} < 10.7$.  

Given this random sampling, simulated galaxies have the average environment and other halo-correlated properties that follow the realized distribution for that halo mass.  For instance, these galaxies will typically have similar values for the halo spin parameter, which has a highly peaked distribution, and will not fully span the possible galaxy diversity owing thereto.  

\begin{figure*}
\begin{center}
\includegraphics[width=6.3in]{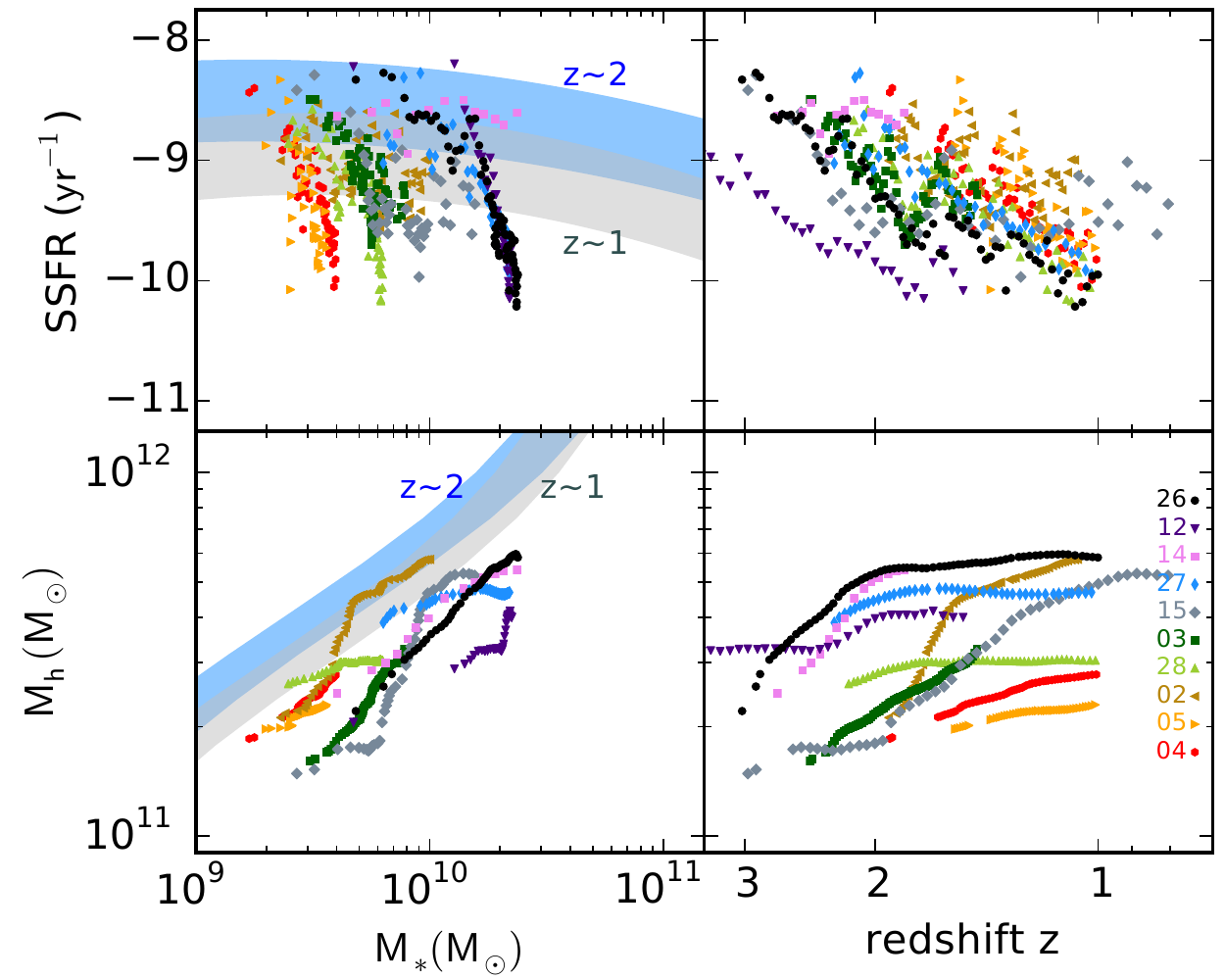}
\caption{  Global properties of the RP simulations studied in this paper:  total stellar mass, halo mass, star formation rate, and redshift.  Point colors correspond with Table~\ref{tab:sims}.  In the top left panel, we show the observed $z=1$--$1.5$ (gray) and $z=2$--$2.5$ (blue) relations for star-forming galaxies from \citet{Whitaker2014} assuming uniform $0.34$ dex scatter.  In the lower left panel, we show the derived $z=1$ (gray) and $z=2$ (blue) relations and errors from \citet{Behroozi2013}. \label{fig:globalparams}}
\end{center}
\end{figure*}

We stored the state of these simulations at a large number of cosmic times, with $\Delta a = 0.01$ at $0.5 < z < 4$, and subsequently analyzed a fraction of these, typically $20$--$100$ per simulation, with the RP simulations sampled more finely than the no-RP simulations.  In many cases we find similar results for the RP and no-RP simulations at the same cosmic time, so for clarity we typically plot the RP simulations only and highlight the key differences where appropriate.  Figure~\ref{fig:globalparams} presents the basic global parameters -- redshift, star formation rate, stellar mass, and halo mass -- of the RP simulations studied here.  The no-RP simulations have $\sim 50\%$ higher stellar mass at the same halo mass.  

Figure~\ref{fig:globalparams} shows that the results of this paper are relevant primarily to galaxies which are not fully quenched, since sSFR $ \gtrsim 10^{-10}\rm\ yr^{-1}$.  This owes to the specific subsample of simulations that we analyzed \citep[contrast with, e.g.,][]{Zolotov2014,Ceverino2015}, and also because we study only sources that would have detectable morphologies in the \candels\ survey (Section~\ref{ss:candelization}).  

\begin{table*}
\caption{RP simulations, plot legend, stellar masses, and Petrosian radii $r_p$.  We measure $r_{p}$ from roughly the rest-frame $B$ filter.  \label{tab:sims}}
\begin{center}
\begin{tabular}{ccccc@{\hspace{1cm}}cc}
Name &plot color& plot symbol & $M_*$ at $z\sim 1$ & $M_*$ at $z\sim 2$ & $r_{p} (B)$ at $z\sim 1$ & $r_{p} (B)$ at $z\sim 2$\\
&&$\log_{10} M_*/M_{\odot}$ & $\log_{10} M_*/M_{\odot}$ & kpc & kpc \\
\hline
VELA04RP & red &hexagon&9.6&9.2&6.8&not measured ($H > 24.5$)\\
VELA05RP & orange &right triangle&9.6&9.3&4.8&not measured ($H > 24.5$)\\
VELA02RP & brown &left triangle&10.0&9.4&6.1&4.8 \\
VELA28RP & green &up triangle&9.8&9.5&12.7&11.6 \\
VELA03RP & dark green &square&9.9&9.7&11.5&6.0 \\
VELA15RP & gray &diamond&10.1&9.8&9.8&6.6 \\
VELA27RP & blue &thin diamond&10.3&10.0&12.0&10.5 \\
VELA14RP & pink &square&10.5&10.2&6.6&6.0 \\
VELA12RP & purple &down triangle&10.4&10.3&10.6&7.3 \\
VELA26RP & black &circle&10.4&10.3&13.5&6.0 \\
\hline
\end{tabular}
\end{center}
\end{table*}

\subsection{Radiative Transfer Calculations} \label{ss:sunrise}

We post-processed the galaxy simulation data into raw mock images using the dust radiative transfer (RT) code \sunrise\footnotemark \citep{jonsson06,jonsson09,Jonsson:2010gpu}, initialized with a pipeline based on the simulation analysis tool \yt\ \citep{Turk2011}.  We first assign to each star particle a spectral energy distribution (SED) based on its mass, age, and metallicity, set by the {\sc Starburst99} stellar population models \citep{leitherer99} with a \citet{Kroupa:2001} initial mass function.  This SED is assigned as a source emitting uniformly from the nearest adaptive refinement tree cell.  For full details, see \citet{Moody2014}.  

Then we calculate the 3-D dust density by assuming it is directly proportional to the metal density predicted by the \art\ simulations.  We assume a dust-to-metals mass ratio of $0.4$ \citep[e.g.,][]{dwek98,James:2002}, and the dust grain size distribution with R=3.1 from \citet{wd01} updated by \citet{draine07}.  This model approximates the average dust observed in the Milky Way \citep[e.g.,][]{Cardelli1989,Gordon2003}.  For stars formed within the last $10^7\rm yr$, we assume an HII + photodissociation region (PDR) model by \citet{groves08} with a PDR covering fraction of 0.2.  This uncertain factor leads to total SEDs matching many normal local galaxies \citep{jonsson09}, but may underestimate the obscuration of massive gas-rich starbursts \citep[e.g.,][]{narayanan10_smg}.  With our focus on galaxies in $M \sim 10^{11.5} M_{\odot}$ haloes at $1 < z < 3$, with few major mergers, we expect the factor 0.2 is a reasonable choice.  Even with the relatively small $17$--$35\rm\ pc$ scales resolved by these calculations, the dust distribution on smaller scales -- which we assume to be uniform -- remains uncertain.

\footnotetext{ \sunrise\ is freely available at http://code.google.com/p/sunrise. }

\sunrise\ performs dust RT using an efficient parallel polychromatic Monte Carlo ray-tracing technique.  Sources emit rays representing their SED, and as each ray propagates through the ISM and encounters dust mass, its energy is probabalistically absorbed or scattered until it exits the grid or enters the pre-defined viewing apertures (``cameras'').   The output of the RT simulation is the SED at each of 600$\times$600 pixels in each camera.  We set 10 cameras, six of which we analyze in this paper.  Of these six, two cameras are aligned to be edge-on and face-on to the angular momentum vector of each galaxy, respectively, and the remaining four are aligned randomly.  For this work, we do not use the dust temperature and emission functionality of \sunrise.  

From these data cubes, \sunrise\ creates raw mock images by integrating the SED in each pixel over the spectral response functions of a set of common astronomical filters, from the far-UV through IR.  We perform this filter synthesis twice: first, we assume the source is at rest with respect to the observer, and second, we assume the source has a cosmological Doppler shift corresponding to the redshift at the cosmic time of the simulation (neglecting peculiar motions).  We create analagous dust-free images in addition to the fully realized images.  The spatial extent of each image increases with mass and cosmic time, and therefore so does the physical pixel scale of the raw mock images.  We use pixel sizes of roughly $50$--$300\rm\ pc$, sufficiently small to robustly simulate \hst\ images of sources at $z \gtrsim 0.2$.

\begin{figure*}
\begin{center}
\includegraphics[width=6.5in]{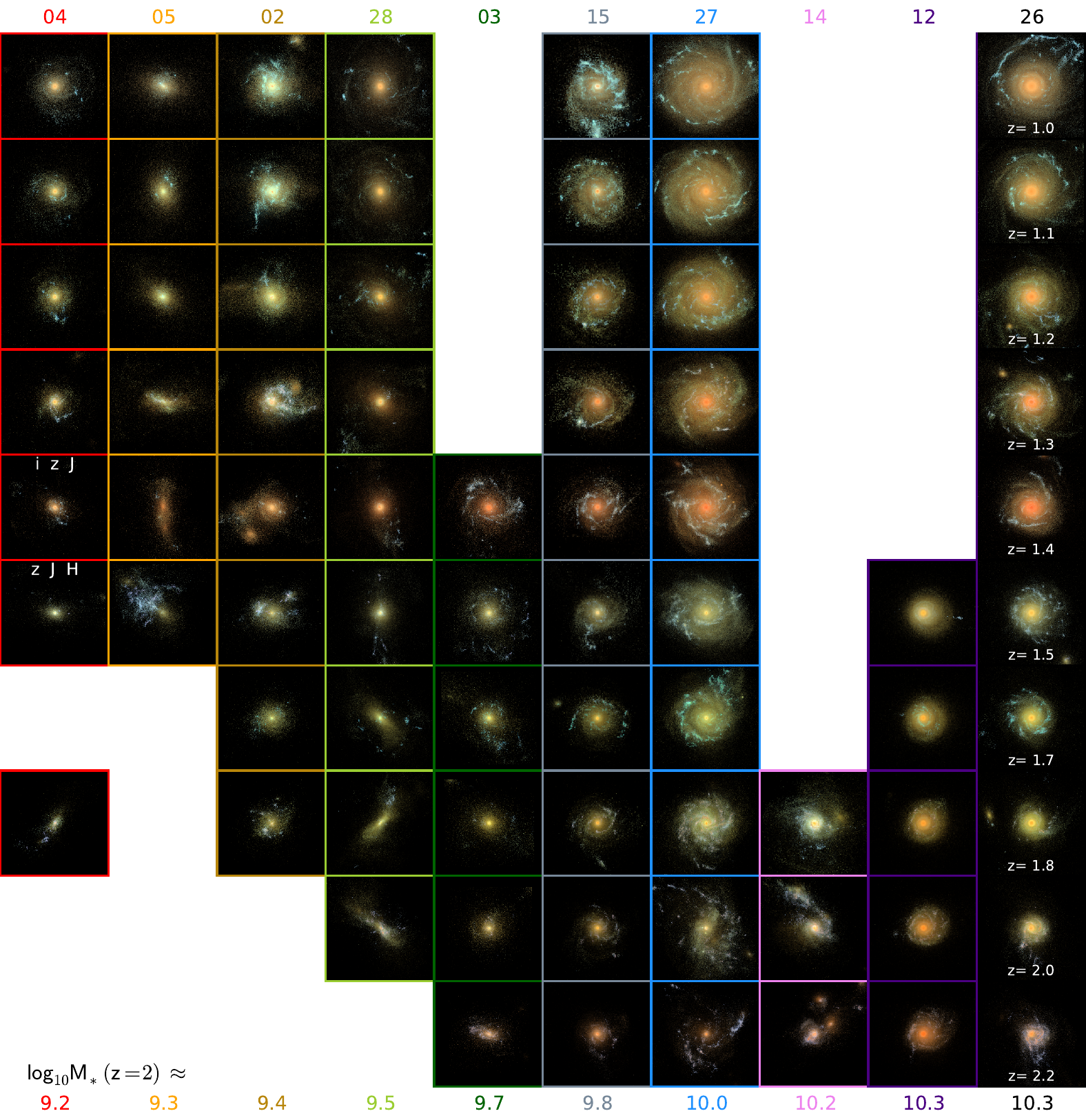}
\caption{  Noiseless high-resolution face-on images $\sim 40$ kpc across for the simulations with radiation pressure (RP), from $z\sim 2$ (bottom) to $z\sim 1$ (top) equally spaced in time.  This shows approximately $20\%$ of the simulation snapshots available, which are stored every $\sim 30$ Myr.  Each column shows the evolution of a single galaxy; the border color corresponds to the points in Figure~\ref{fig:globalparams} and Table~\ref{tab:sims}.  We derive the color and intensity at each pixel from synthetic mock images calculated with \sunrise, using the scaling algorithm by \citet{Lupton2004}.  To create roughly rest-frame U-B-V images, at $z > 1.5$ we show the z, J, and H filters while at $z < 1.5$ we show i, z, and J.  Here we use the full pixel resolution from the simulated images, with no noise or PSF convolution. At low redshifts, some columns have missing data because the simulations were not run past that point.  At high redshifts, some columns have missing data because a source was either not detected at all or not analyzed ($H > 24.5$) in our photometry and morphology pipeline from the \candels-like images (Figure~\ref{fig:hstimages}). \label{fig:simimages}}
\end{center}
\end{figure*}

\subsection{Image Realism} \label{ss:candelization}

We include a number of \hst\ UV through IR broadband filters in our \sunrise\ pipeline.  For this work, we select five filters to analyze, focusing on simulating a subset of the filters used by the \candels\ Multi-Cycle Treasury Project (\citealt{Koekemoer2011}, \citealt{Grogin2011}).   They are:
\begin{align*}
&\textrm{V\quad (ACS/F606W)} \\
&\textrm{I\quad (ACS/F775W)}\\
&\textrm{Z\quad (ACS/F850LP)}\\
&\textrm{J\quad (WFC3/F125W), and}\\
&\textrm{H\quad (WFC3/F160W).} 
\end{align*}

In the same manner as the \candels\ survey, this permits us to study, at high spatial resolution, a consistent rest-frame optical band across roughly $5\times10^9\rm\ yr$ at $0.5 < z < 3$, mitigating the significant and confounding effect of the wavelength dependence of galaxy morphology.  Additional filters are available in the \sunrise\ pipeline for future analyses. In Figure~\ref{fig:simimages} we show high-resolution noiseless simulated images of the ten RP simulations on which we focus, from $z\sim 2.3$ to $z\sim 1$.  To create roughly rest-frame U-B-V images, at $z > 1.5$ we show the z, J, and H filters while at $z < 1.5$ we show i, z, and J.  

\begin{figure*}
\begin{center}
\includegraphics[width=6.5in]{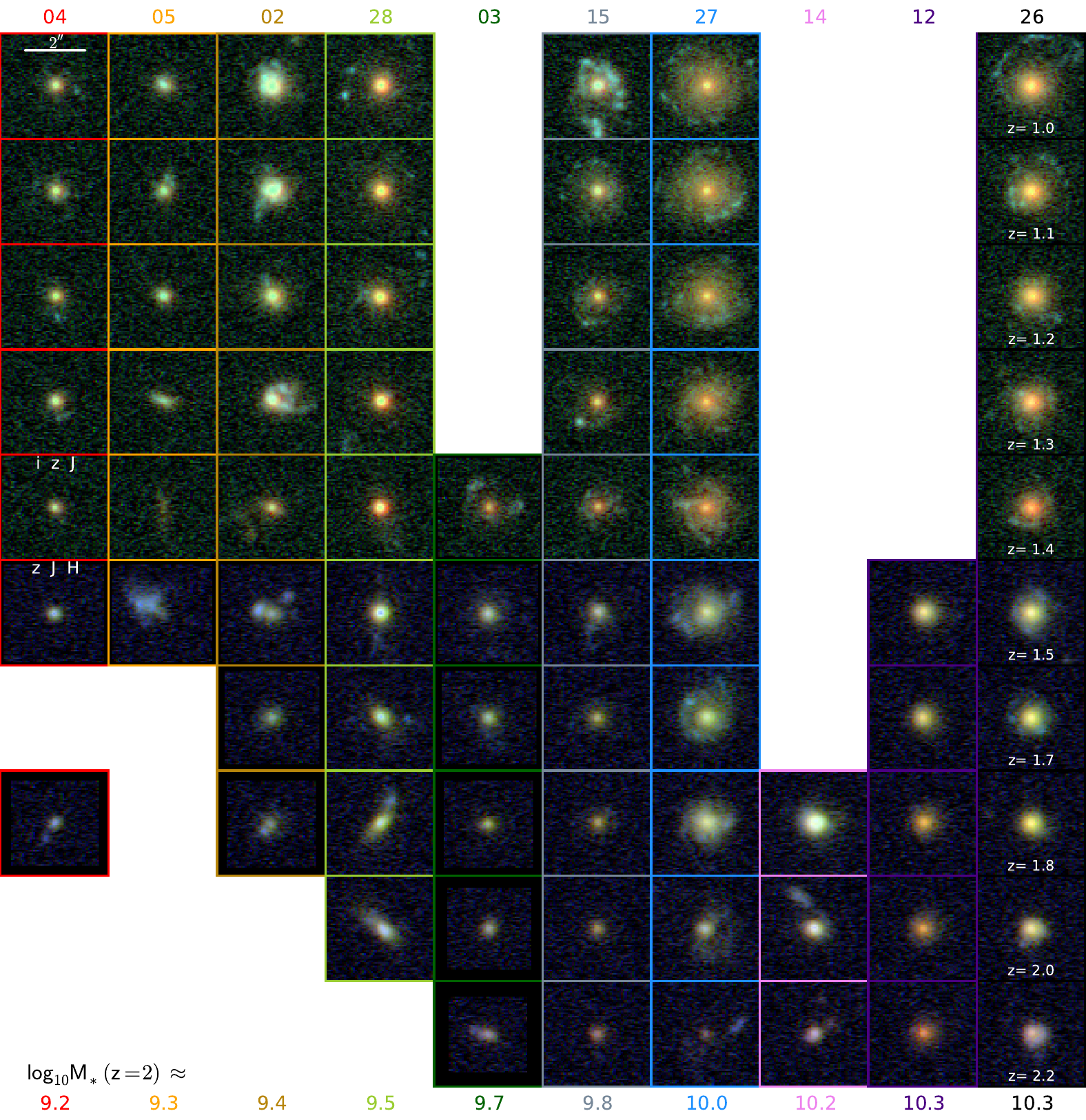}
\caption{ Same as Figure~\ref{fig:simimages} except showing realistic mock \hst\ images.   Each panel is $5.4\rm\ arcsec$ ($\sim 40$--$45$ kpc) across, and arranged from $z\sim 2$ (bottom) to $z\sim 1$ (top) equally spaced in time.  Each column shows the evolution of a single galaxy; the border color corresponds to the points in Figure~\ref{fig:globalparams} and Table~\ref{tab:sims}.  To generate these, we smoothed the images from Figure~\ref{fig:simimages} with appropriate PSFs, binned them to a pixel scale of $0.06\rm\ arcsec$ equal to the \candels\ high-level science images \citep{Koekemoer2011}, and added a random background at the level appropriate for each filter \citep{Grogin2011}.  The result is a direct prediction for how these models would appear in \candels, allowing us to evaluate possible differences between observed stellar structures and galaxy formation theory.  At low redshifts, some columns have missing data because the simulations were not run past that point. At high redshifts, some columns have missing data because a source was either not detected at all or not analyzed ($H > 24.5$) by our photometry and morphology pipeline based on these \candels-like images.\label{fig:hstimages}}
\end{center}
\end{figure*}

From these, we scale the raw pixel sizes according to the angular size distance and fluxes for the luminosity distance at the redshift of each simulated source.   Then we convolve these raw mock \hst\ images with model point-spread functions (PSFs) appropriate for each instrument/filter combination \citep{Krist2011}, bin them to a pixel scale of $0.06\rm\ arcsec$, and add noise approximating the total random background of the \candels-Wide survey in each filter \citep{Grogin2011}.  The result is a database of model images at the resolution and depth of \candels-Wide. This technique follows \citet{lotz08}, and is similar to that used by \citet{Wellons2014} for creating mock \hst\ images, and by \citet{Torrey2015} to create mock SDSS-like images. In Figure~\ref{fig:hstimages} we show the final synthetic \hst\ images in the same filters and spatial scale as in Figure~\ref{fig:simimages}.

We can adjust at will the depth of the mock images, and in Section~\ref{ss:depth} we discuss a case where we required more depth to obtain reliable optical morphologies from the simulation (VELA28MRP; light green points).

\subsection{Non-parametric Diagnostics}   \label{ss:pipeline}

We locate and analyze sources in these images following closely the techniques used by \candels.  We apply {\sc SExtractor} \citep{bertin96} in dual-image mode to PSF-matched images, detecting sources in H, and use a multi-step algorithm to robustly locate small, faint sources while maintaining the contiguity of larger, brighter galaxies \citep{Galametz2013}.  The result is a list of positions, sizes, and photometry for zero or more individually detected sources in each image.   In this paper, we analyze the sources associated with the halo targeting described in Section~\ref{ss:hydroart}, which we accomplish in practice by focusing on the bright source centered in each image.  

For all such sources with $H < 24.5$, we measure non-parametric structural parameters in all five filters, including the Petrosian radius $r_P$, half-light radius $R_{1/2}$, Concentration (C) and Asymmetry (A) as defined by \citet{Conselice2003a}, Gini and $M_{20}$ as defined by \citet{Lotz2004}, as well as three new indicators Multimode (M), Intensity (I) and Deviation (D) statistics by \citet{Freeman2013}.  These quantities describe the light profile of observed sources in a computationally efficient manner, and have been used to classify galaxy structural types and morphological disturbances in numerous surveys, including \candels\ (Peth et al.\ in prep.).  \citet{Grogin2011} showed that the $H < 24.5$ magnitude selection cut is effective at identifying galaxies that are detected well enough to measure morphology in both non-parametric diagnostics \citep[e.g.,]{Lotz2011} and visual classifications \citep{Kartaltepe2014}.  

We use code originally developed for idealized merger simulations \citep{lotz08,lotz10,Lotz2010} and also applied to galaxy surveys \citep{Lotz2004,lotz08_hst,Lotz2011}. We will first focus on the basic structural evolution of the cosmological simulations, but we return briefly to merger diagnostics in Section~\ref{ss:mid}.

The Petrosian radius $r_P$ is defined such that the mean surface brightness in an elliptical annulus at $r_P$ equals 0.2 times the mean surface brightness within $r_P$. We define a galaxy's pixels as those with flux values greater than the mean surface brightness at $r_P$.  

We use the concentration parameter $C$ \citep{Bershady2000}:
\begin{equation}
C = 5\log_{10} \frac{r_{80}}{r_{20}},
\end{equation}
where $r_{80}$ and $r_{20}$ are circular apertures containing $80\%$ and $20\%$ of the total flux within $1.5 r_P$ \citep{Conselice2003} of the galaxy centre defined by minimizing the Asymmetry parameter \citep{Abraham1996}.

Gini's coefficient ($G$) measures inequality among a galaxy's pixel flux values, varying from $0$ (all pixels equal) to $1$ (one pixel contains all flux).   \citet{Abraham2003} first used it to characterize galaxy light profiles.  $G$ increases with $C$ but does not depend on the location of the brightest pixels.  Hence it detects not only compact spheroids but also galaxies with multiple cores.  For a discrete sample, \citet{Glasser1962} showed that $G$ can be computed as:
\begin{equation}
G = \frac{1}{ \bar{\left | I_i\right |} n\left (n-1\right )} \sum_i^n{\left (2i - n - 1\right ) \left | I_i\right | },
\end{equation}
where we have $n$ pixels with rank-ordered absolute flux values $\left | I_i\right |$, and $\bar{\left | I_i\right |} = \sum_i{\left | I_i\right |/n}$, the mean absolute flux value.  We follow \citet{Lotz2004} in correcting $G$ using absolute values to mitigate the effect of noise-induced negative fluxes.  This procedure recovers the true $G$ when $S/N \gtrsim 3$ per galaxy pixel, which is true for almost all galaxies with $H < 24.5$ (see Section~\ref{ss:depth} for a counterexample).  

$M_{20}$ is the spatial moment of a galaxy's brightest quintile of pixel flux values, relative to its total moment \citep{Lotz2004}.  
\begin{equation}
M_{20} \equiv \log_{10} \frac{\sum_i{M_i}}{M_{\rm tot}},\ \mathrm{for}\ \sum_i{I_i} < 0.2 I_{\rm tot},
\end{equation}
where
\begin{equation}
M_{\rm tot} = \sum_i^n{M_i} = \sum_i^n{I_i \left [  \left (x_i - x_c\right )^2 + \left (y_i - y_c\right )^2 \right ]},
\end{equation}
and $x_c$, $y_c$ are the 2-D spatial coordinates of the galaxy centre, defined to minimize $M_{\rm tot}$.  

{ These structural diagnostics correlate with \sersic\ index $n_S$, and we show this correlation for the present simulations in Section~\ref{ss:sersic}.  Observationally, $n_S$, $M_{20}$, $G$, and $C$ all trace the strength of the bulge component in the light profile.  However, numerically, \gmtwenty\ tend to spread out disc-dominated galaxies across many values and concentrate bulge-dominated ones, while the opposite is true for $n_S$.  Combinations of non-parametric diagnostics have been shown to correlate better with quenching than $n_S$ alone (Peth et al.\ in prep.).  Moreover, in low-redshift galaxies, these diagnostics correlate in the expected ways with galaxies' stellar kinematics, albeit with large scatter (Snyder et al.\ in prep.).  }

The $M$, $I$, and $D$ statistics were introduced by \cite{Freeman2013}
to automatically identify
disturbed morphologies in a way that reproduces the results of visual
classifications. 
Let $S_l$ be a superlevel set for $I$, i.e.,
$S_l$ is the collection of pixels within a galaxy's segmentation map with intensity greater than or equal
to a given threshold $l$. Given this collection, one
groups all contiguous pixels, orders the groups by decreasing area (such that
$A_{l,(i)}$ is the area of the $i^{\rm th}$ largest group), and sets
\begin{eqnarray}
R_l = \frac{A_{l,(2)}}{A_{l,(1)}} A_{l,(2)} \,. \label{eqn:ar}
\end{eqnarray}
The $M$ statistic is then the maximum $R_l$ value:
\begin{eqnarray}
M = \max_l \, R_l \,. \label{eqn:mstat}
\end{eqnarray}
$M$, defined in this fashion, is not dimensionless, and thus a galaxy's 
$M$ value can depend on its angular size distance. While we choose to use
$M$ as defined above in this work, we will consider alternative definitions
for future analyses, such as
\begin{eqnarray}
R_l = \frac{A_{l,(2)}}{A_{l,(1)}} \frac{A_{l,(2)}}{n} \,, \label{eqn:ar2}
\end{eqnarray}
where $n$ is the number of pixels within the galaxy's segmentation map, or
\begin{eqnarray}
R_l = \frac{A_{l,(2)}}{A_{l,(1)}} \,. \label{eqn:ar3}
\end{eqnarray}
The normalization in Equation~\ref{eqn:ar2} fixes the possible range of $M$ to the interval $[0,0.5)$,
where 0.5 would correspond to the situation, impossible to observe in practice,
for which $A_{l,(1)} = A_{l,(2)} = n/2$.  The normalization in Equation~\ref{eqn:ar3} fixes the possible range of $M$ to the interval $[0,1)$.

The $I$ statistic is complimentary to $M$ in that it takes into account
pixel intensities. To compute $I$, one traces maximum gradient paths from each
pixel within a galaxy's segmentation map to corresponding local intensity 
maxima, i.e.,
for a given starting pixel, one examines the eight surrounding pixels and moves
to the pixel offering the largest increase in intensity, and repeats the
process until a local maximum is reached. Each local maximum is thus
associated with a set of pixels $p$ with summed intensity $I_p$. 
One sorts the summed intensities for each pixel set in decreasing order, 
yielding the values $\{I_{(1)},I_{(2)},\cdots\}$. The $I$ statistic is then
\begin{eqnarray}
I = \frac{I_{(2)}}{I_{(1)}} \,. \label{eqn:istat}
\end{eqnarray}
Note that in practice, one obtains more accurate morphological 
classifications with $I$ if the image data are slightly smoothed before
$I$ is computed. For instance, \cite{Freeman2013}, who worked with HST images
from the GOODS-S field, smoothed their image data with a symmetric
Gaussian kernel with $\sigma \approx$ 1 pixel.

The $D$ statistic identifies galaxies whose shapes deviate from elliptical
symmetry, and thus it serves a similar function to the $A$ statistic defined
above. One defines the intensity centroid of a galaxy as
\begin{eqnarray}
(x_{\rm cen},y_{\rm cen}) = \left( \frac{1}{n} \sum_{i=1}^n{x_i I_i} , \frac{1}{n} \sum_{i=1}^n{y_i I_i} \right) \,, \nonumber
\end{eqnarray}
with the summation being over all pixels within the galaxy's segmentation map.
The $D$ statistic is then
\begin{eqnarray}
D = \sqrt{\frac{\pi}{n}} \sqrt{(x_{\rm cen}-x_{I_{(1)}})^2 + (y_{\rm cen}-y_{I_{(1)}})^2} \,, \label{eqn:dstat}
\end{eqnarray}
where $x_{I_{(1)}},y_{I_{(1)}}$ is the local maximum corresponding
to the set of pixels with summed intensity $I_{(1)}$, and
$\sqrt{n}/\pi$ is an approximate galaxy ``radius" that acts to
normalize $D$.

In Section~\ref{s:morphs}, we present the basic structural evolution of the simulated galaxies as if they were observed in \candels.


\begin{figure}
\begin{center}
\includegraphics[width=3.0in]{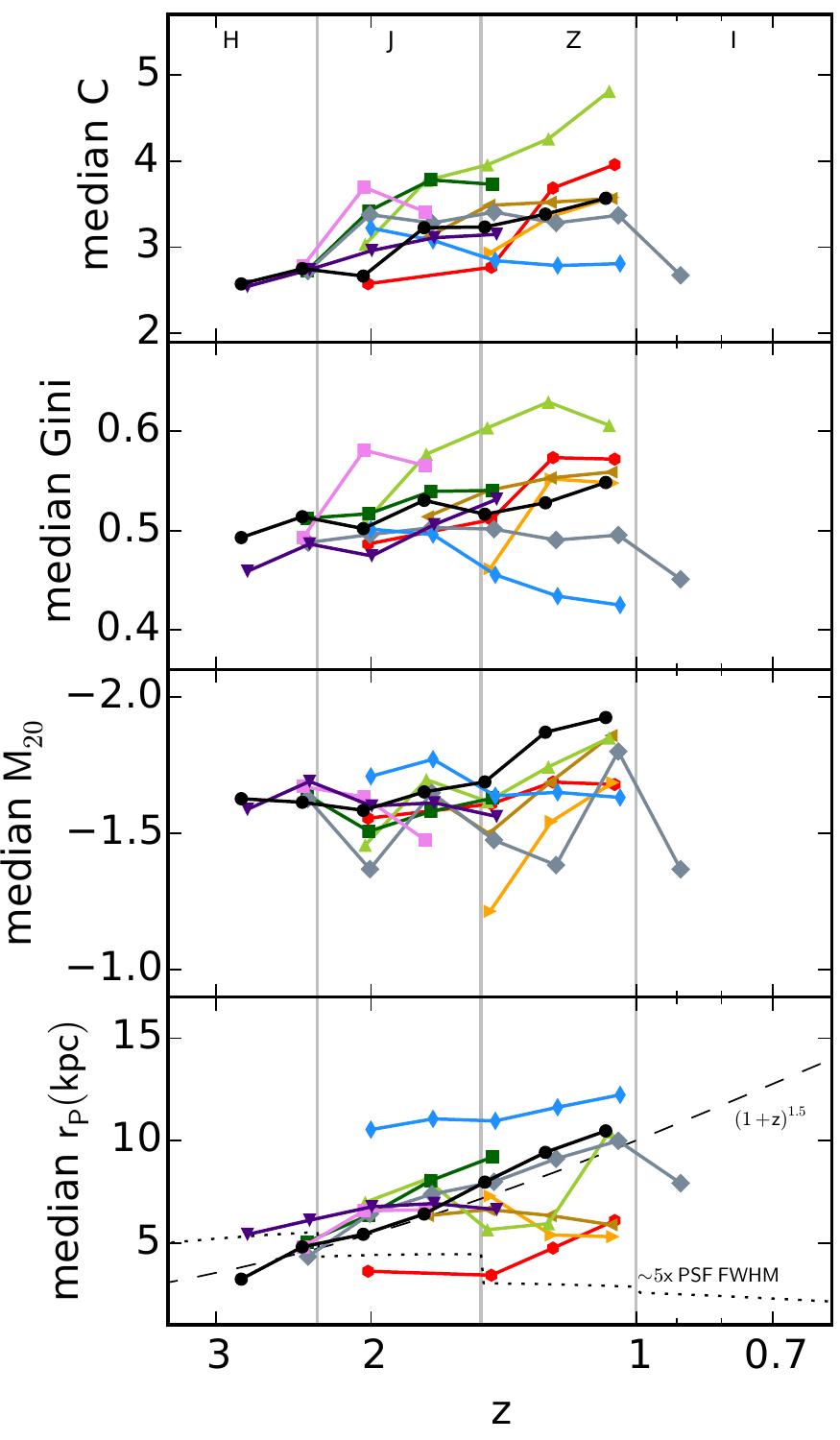}
\caption{ Median of binned structural parameters Gini, $M_{20}$, and Petrosian radius $r_P$ versus redshift for the 10 RP simulations studied in this work: colors correspond with Table~\ref{tab:sims} and Figure~\ref{fig:globalparams}.  We measure each quantity in roughly the same rest-frame B filter, corresponding to the observed-frame filters labelled at the top of the first panel and delimited by the light gray vertical lines.   \label{fig:structureparams}}
\end{center}
\end{figure}

\section{Morphology Evolution}  \label{s:morphs}

In this paper, we explore the predicted evolution of basic galaxy morphology.  Therefore we analyze primarily diagnostics of the overall light profile: $R_{p}$, G, $M_{20}$, and C.  In this section we present the time-evolution of these parameters in the ten VELARP simulations, which have $9.2 < \log_{10} M_*/M_{\odot} < 10.3$.  For this purpose, at a given redshift we focus on the \hst\ filter corresponding most closely to the rest-frame B filter centered at $\sim 0.45\mum$ \citep{johnsonmorgan53}:  
\begin{align*}
&z > 2.3\quad\quad \textrm{H band} \quad\quad \sim 1.6\mum \\ 
1.5 \le\ &z < 2.3 \quad\quad \textrm{J band} \quad\quad \sim 1.25\mum\\ 
1.0 \le\ &z < 1.5 \quad\quad \textrm{Z band} \quad\quad \sim 0.85 \mum\\
0.6 \le\ &z < 1.0 \quad\quad \textrm{I band} \quad\quad \sim 0.78\mum\\ 
&z < 0.6\quad\quad \textrm{V band} \quad\quad \sim 0.61\mum.
\end{align*}
Each simulation consists of $\sim 200$ timesteps at which we conducted the radiative transfer post-processing.  Of these, $\sim 50$ per simulation contained a source bright enough to be detected and its morphology measured from the mock images in at least one of the four randomly oriented cameras.  We defined our nominal detection threshold in Section~\ref{ss:pipeline}.  Unless otherwise noted, we derive results only from this set.

\subsection{Visual morphology}

In Figure~\ref{fig:hstimages} we can directly observe some basic trends.  Galaxies grow from bottom to top, where each column represents one of the ten RP simulations, organized from left to right by increasing stellar mass at $z=2$ as in Table~\ref{tab:sims}.  At $z\sim2$, most objects are compact, potentially hosting unresolved star-forming discs.  By $z\sim1.5$ these simulations host red compact bulges, and both disc and bulge components tend to grow between $z=1.5$ and $z=1$.  At $z \lesssim 1.5$ we see extended star-forming discs around roughly half of the set.  At $z\sim2$, three simulations with $M_* < 10^{9.5} M_{\odot}$ are either undetected or do not satisfy our $H < 24.5$ selection criterion.  

The incidence of large star-forming discs at $z < 1.5$ (top half) in the color images of Figure~\ref{fig:hstimages} appears to increase with stellar mass.  We recover this trend quantitatively in the subsequent figures: galaxies in our sample with $M_* < 10^{10} M_{\odot}$ tend to be compact bulges at $z < 2$, while the higher mass model galaxy population tends to contain both disc-dominated and bulge-dominated objects.  This trend may be expected as a consequence of increasing star formation efficiency as one approaches the peak of the stellar mass-to-halo mass relation from below.  It may also reflect ``disc settling'' seen at $z < 1$ \citep{Kassin2012} whereby more massive galaxies tend to have more ordered motions typical of rotating discs.  

{ In the high-resolution images of Figure~\ref{fig:simimages}, we can get a clearer understanding of the galaxies' evolutionary history. At several timesteps, simulated galaxies experience rapid or violent phenomena such as mergers, disc formation, and clumpy star formation.  At many other times, such as the later stages of VELA04MRP, VELA27MRP, and VELA28MRP, the galaxies form in a smooth or slow fashion, such that their high-resolution optical images often resemble spiral galaxies in the low-redshift universe. }

\subsection{As a function of time}

\begin{figure*}
\begin{center}
\includegraphics[width=5.5in]{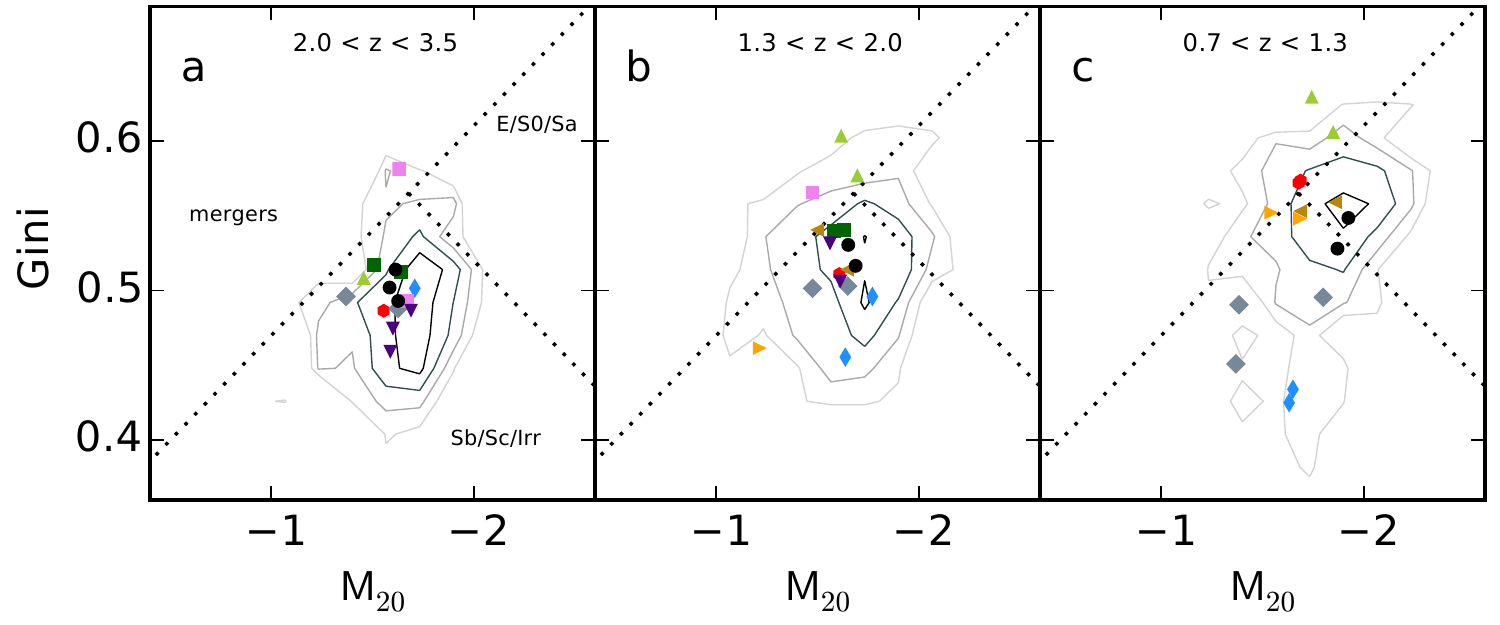} 
\caption{Top:  Gini--$M_{20}$ diagram versus redshift for the RP simulations, in three panels: a) $z > 2$, b) $z \sim 1.5$, c) $z \sim 1.0$.   Darker contours enclose regions of increasing logarithm of relative number density, evenly spaced.  This includes all simulated timesteps considered in this work, for which individual simulations may be weighted unevenly.  The points are the median binned values presented in Figure~\ref{fig:structureparams}, evenly spaced in redshift, and therefore individual simulations are weighted equally insofar as data exists at a given epoch.  Outliers in Gini from Figure~\ref{fig:structureparams}, the light green points of panel b fall in the ``merger'' area of this simple classification diagram. However, this simulation (VELA28RP) has an extremely low-surface-brightness extended disc that is not apparent in the \candels-wide images. At $z\sim 2$--$1.5$, its compact core but numerous faint pixels lead to the large Gini values.  See Section~\ref{ss:depth} for further discussion on this simulation. \label{fig:versus_z}}
\end{center}
\end{figure*}

In Figure~\ref{fig:structureparams} we show how $R_{p}$, $G$, $M_{20}$, and $C$ evolve with time in each of the ten RP models, at roughly the rest-frame B band as described above.  For each bin evenly spaced in redshift, we plot the median value of the parameters measured within that bin.  From these average quantities, the basic result is that the morphological evolution is not uniform.  While these models are confined to a fairly small range in these quantities at $z \gg 2$, they evolve along diverse paths such that the distribution expands through $z \sim 1.5$.  There are two simulations clustered at $G\sim0.45$ ($C\sim 3$), corresponding to a relatively discy light profile, while the rest are more bulge-like, having median $G \gtrsim 0.53$ ($C \gtrsim 3.5$).  This quantitative separation corresponds well with the visual impressions in Figure~\ref{fig:hstimages}.  All 10 RP simulations have similar median $M_{20}$ values during these times.

Size evolution of the simulated galaxies is similarly diverse.  The bottom panel of Figure~\ref{fig:structureparams} shows that their radii ($r_P$) evolves roughly proportional to $(1+z)^{1.5}$, with a wide scatter.  Some galaxies (pink, orange, brown) have roughly constant median sizes at $1 <z < 3$, while others grow dramatically (light green; see subsequent figures and Section~\ref{ss:mergerdiskgrowth}).  Depending on their mass, the ten RP simulations have $r_P$ larger than $\sim 5\times$ the PSF in the measured filter by $z \sim 2.5$.  We also estimated half-light radii, which are roughly $\sim 2-4$ kpc among the simulations we studied at $z=1$-$2$.  These are measured from roughly the rest-frame $B$ filter and do not correct for PSF, and so may not accurately trace the 3D half-mass radii of these galaxies, which \citet{Zolotov2014} measured to be a factor of 2-4 smaller in several cases than those we measured here.

In Figure~\ref{fig:versus_z} we view the simulated galaxies' evolution in \gmtwenty\ space. We plot the same median values as colored points in three redshift bins.  These points are evenly spaced in redshift, and therefore at a given epoch each simulation has equal weight.  We plot dotted lines that coarsely separate observed galaxies into late types (bottom triangle), early types (right triangle), and mergers (upper left segment) following \citet{Lotz2004}.  In addition, we compile all available models at each simulated timestep and plot them in contours of relative number density.  This allows us to see how these values evolve in multiple dimensions: as in Figure~\ref{fig:structureparams}, the median values split into a clearly wider distribution at $z < 1.3$, while the overall average trend with time is from the centre of the diagram upward and to the right, toward higher $G$ and lower $M_{20}$ (more bulge-like).  

\subsection{As a function of mass}

\begin{figure}
\begin{center}
\includegraphics[width=3.2in]{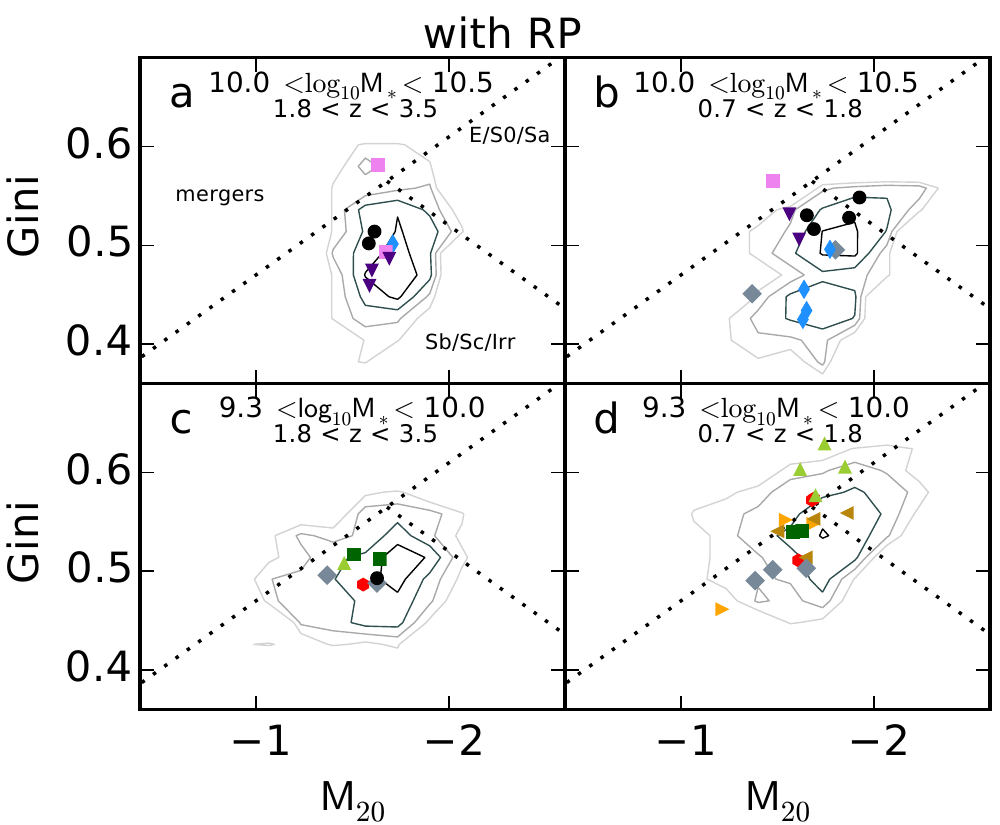}
\includegraphics[width=3.2in]{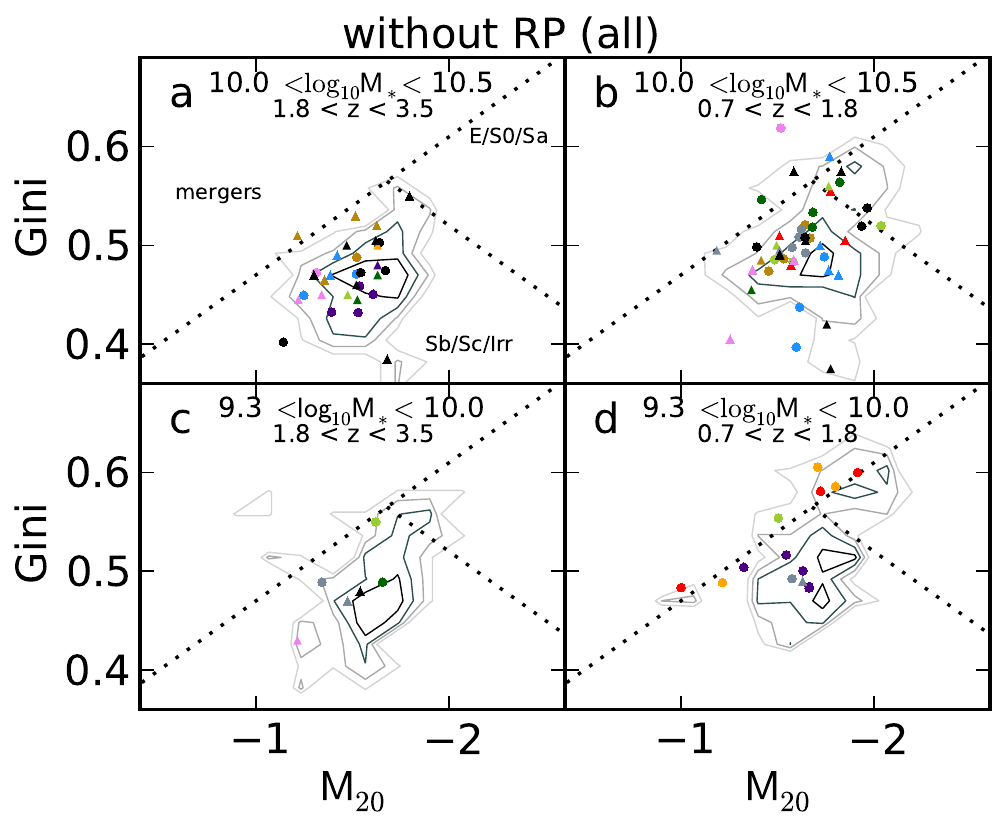}
\caption{ Gini--$M_{20}$ versus stellar mass and redshift, in four panels: a) high mass, high redshift b) high mass, low redshift, c) low mass, high redshift, d) low mass, low redshift.  Contours and points have the same meaning as in Figure~\ref{fig:versus_z}, where colored circles correspond to the same haloes simulated in both the RP and no-RP simulations: colored triangles are the no-RP simulations that do not presently have counterparts with synthetic images.  In both the RP and no-RP simulations, there are relatively more disc-dominated galaxies in the higher mass bin at $z < 1.8$ (right columns). Moreover, the overall distribution of points is similar in both the RP and no-RP simulations.  This suggests that the kpc-scale morphological features probed by \gmtwenty\ are somewhat insensitive to the detailed implementation of feedback from supernovae and massive stars.  Instead, these features are likely governed primarily by the galaxy's gas accretion and assembly history.  \label{fig:mass_and_z}}
\end{center}
\end{figure}

In Figure~\ref{fig:mass_and_z}, we separate these model data into four quadrants at $M_* = 10^{10} M_{\odot}$ and $z = 1.8$.  Since a galaxy's structure and star formation rate correlate with its mass in previous studies at these redshifts \citep[e.g.,][]{Wuyts2011,Lee2013}, we might expect to see a trend in our model structural diagnostics as a function of mass.  

In the high redshift bin (left column), points and contours are centered in the diagram for both mass bins.  This implies that the vast majority of model galaxies have disc-dominated profiles.  Contours in the right column show that the overall trend toward the early-type region is present to roughly the same extent in both mass bins: some bulge growth occurs at all masses.  Almost none of our models occupy the most extreme elliptical end of the \gmtwenty\ locus (having $G \sim 0.6$, $M_{20} \sim -2.5$).  

On the other hand, the models with $M_* > 10^{10} M_{\odot}$ contain more objects with robustly disc-dominated profiles ($G < 0.5$), and we see that these points (blue and gray) in the upper right panel are what gives rise to the structural diversity by $z \sim 1.3$ in Figures~\ref{fig:structureparams} and \ref{fig:versus_z}.

\subsection{As a function of star formation}

In Figure~\ref{fig:ssfr_and_z}, we separate the models by specific star formation rate (SSFR$ = $SFR$/M_*$) and redshift.  We chose a limit of $SSFR=10^{-9.7}\rm\ yr^{-1}$ so that the ``low star formation'' bin contains a significant number of galaxies.  However, this SSFR is a factor of $\sim10$ higher than common definitions of quiescent or non-star-forming levels \citep[e.g.,][]{brammer09}.  Therefore this should be considered a cut that classifies galaxies into bins of normal star formation versus low star formation, as opposed to ``star forming'' and ``not star forming''.  

We find a difference in the average morphologies between the normal and low SF bins at $z < 1.8$ (right column) in Figure~\ref{fig:ssfr_and_z}.  In the models, normal star-formers prefer the late-type (lower centre) region of the \gmtwenty\ diagram, while the low star-formers tend toward the observed position of early-type galaxies \citep{Lotz2004} in the right segment.  One of the simualtions that becomes disc dominated with time (blue; VELA27RP) is again visible as an outlier in the low star star formation, $z < 1.8$ panel.  In general, normal star-forming galaxies are primarily disc-like, while low-star forming galaxies are either disc-like or bulge-like.

\subsection{RP versus no-RP simulations} \label{ss:rpcompare} \label{ss:rptest}

\begin{figure}
\begin{center}
\includegraphics[width=3.2in]{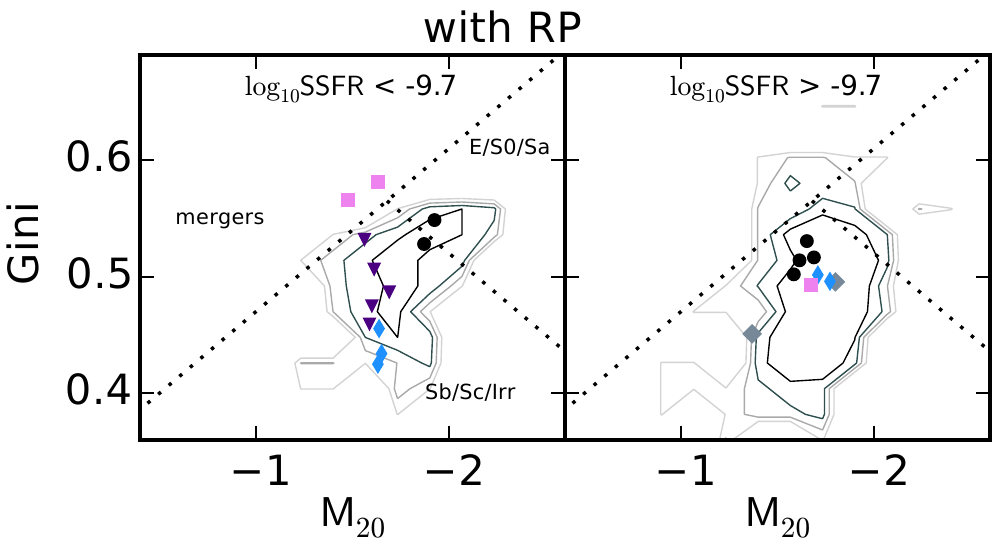}
\includegraphics[width=3.2in]{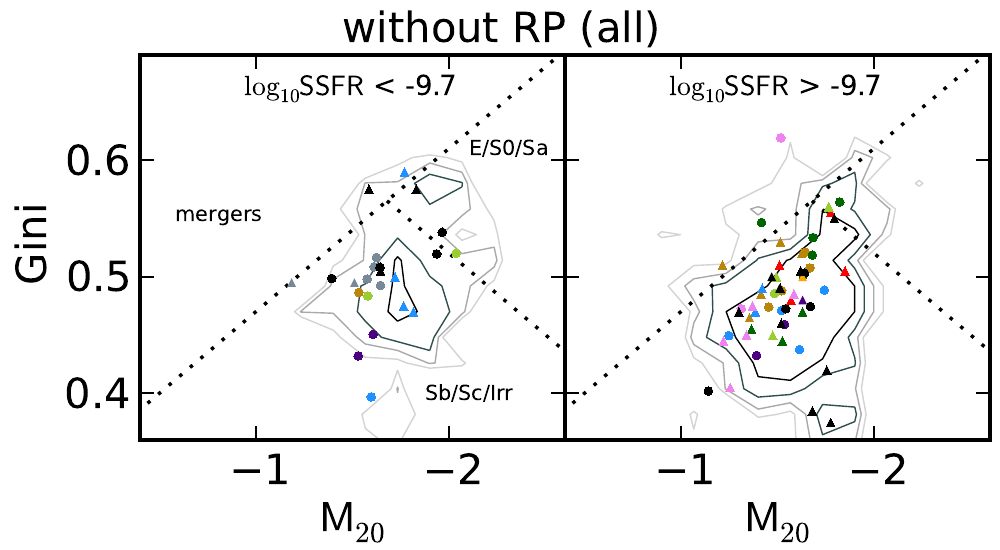}
\caption{\gmtwenty\ versus specific star formation rate (SSFR = $\rm SFR/M_*$) for the same galaxies simulated with and without radiation pressure (RP) feedback, for models with $M_* > 10^{10} M_{\odot}$.   Contours and points have the same meaning as in Figure~\ref{fig:versus_z}.  As in Figure~\ref{fig:mass_and_z}, the distribution of \gmtwenty\ and their trends with SFR and mass are broadly similar in the two cases. \label{fig:ssfr_and_mstar}}
\end{center}
\end{figure}

The phenomena described above are also present in the no-RP simulation set (Figures~\ref{fig:mass_and_z} and \ref{fig:ssfr_and_mstar}), which contains more than twice as many galaxies with synthetic images calculated, and in particular a broader range of masses.  At $z > 1.8$, the no-RP simulations have a somewhat greater fraction of low-mass galaxies that are very disc-dominated ($G \lesssim 0.45$), in contrast to the RP simulations, but relatively fewer than at higher mass.  

\begin{figure}
\begin{center}
\includegraphics[width=3.2in]{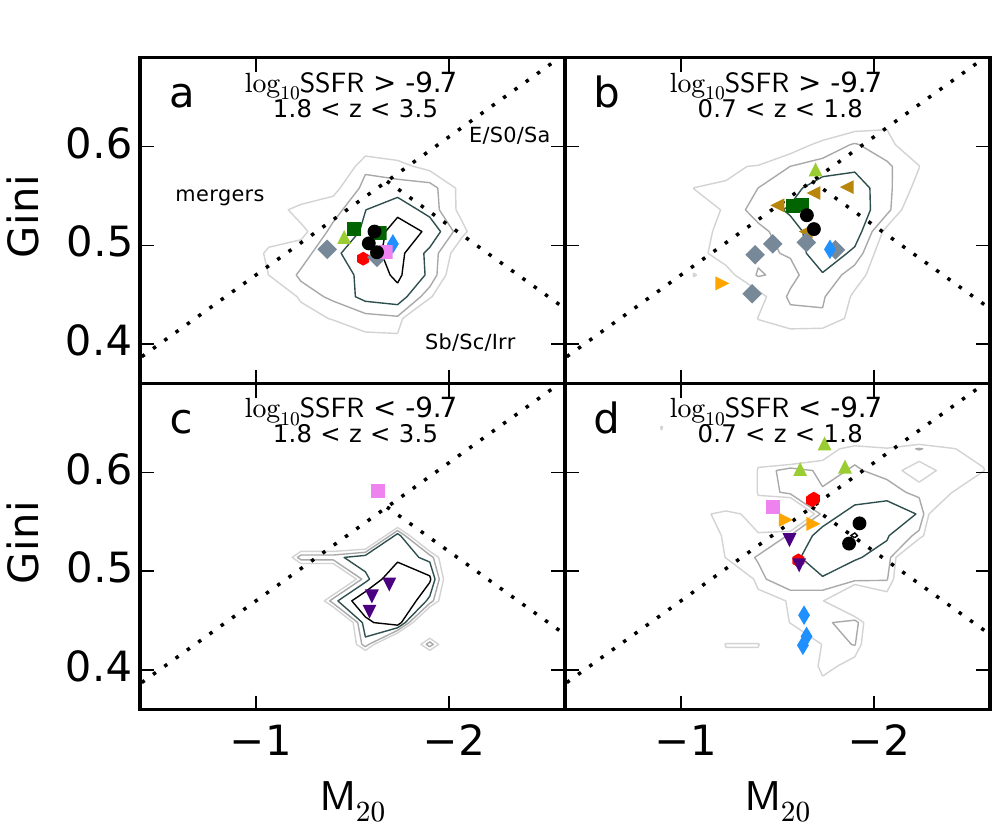}
\caption{Gini--$M_{20}$ versus specific star formation rate (SSFR = $\rm SFR/M_*$) in four panels: a) high SSFR, high redshift b) high SSFR, low redshift, c) low SSFR, high redshift, d) low SSFR, low redshift.   Contours and points have the same meaning as in Figure~\ref{fig:versus_z}.  \label{fig:ssfr_and_z}}
\end{center}
\end{figure}

Overall, the general distribution of points in \gmtwenty, and their dependence on mass and SFR, are very similar under the two different feedback models.  Specifically, Figure~\ref{fig:ssfr_and_z} shows that in both the RP and no-RP simulations, nearly all of the vigorously star-forming simulated galaxies have \gmtwenty\ values in the bottom region of the diagram: they are disc-dominated light profiles. By contrast, the galaxies with $SSFR < 10^{-9.7}\rm\ yr^{-1}$ are spread across the region separating disc-dominated and bulge-dominated.

In these cases of a minor shift or no change in the \gmtwenty\ distributions, it is difficult to infer anything concrete about the effects of RP feedback \citep[c.f.,][]{Trujillo-Gomez2013,Moody2014}.  However, it is important not only to predict cases where observables will differ, but also to highlight certain trends which may be insensitive to differences in the model.  In this case, \gmtwenty\ (and $C$) probe the kpc-scale distribution of observed optical light.  In an individual galaxy, these measurements may differ quantitatively depending on whether the RP or no-RP feedback model is used (e.g., orange circles in upper right panel of Figure~\ref{fig:mass_and_z}).  We have not performed a systematic comparison on a per-galaxy basis.  However, on average, \gmtwenty\ has the same distribution of values in both the RP and no-RP simulations. This does not rule out that other types of measurements may be significantly more sensitive to this difference.

\subsection{Effects of dust}

Dust affects the measured diagnostics in a predictable way.  In particular, we find that the rest-frame B band $G$ values increase by $\sim 0.1$ when the \sunrise\ dust RT is turned off, reflecting increased surface brightness of the few pixels containing young stars.  Here, turning off dust RT corresponds with removing dust from the galaxies' diffuse ISM component -- there remains sub-grid dust obscuration by the PDR around stars with ages $t < 10^7\rm\ yr$.  Given our dust RT settings in Section~\ref{ss:sunrise}, the former component has a significantly larger obscuring effect in the rest-frame optical.  

Since the observed morphology distribution declines sharply at $G > 0.6$ \citep[especially for $M_{20} \gg -2$;][]{lotz08_hst}, we conclude that an accurate treatment of dust attenuation is essential to bring these model diagnostics into agreement with data, similarly found by \citet{lotz08} in qualitatively different simulations.  Removing dust also slightly increases the detectability of the models in our rest-frame optical, \candels-like mock images.  

Without diffuse dust attenuation, the emergence of a connection between morphology, mass, and star formation in these models remains qualitatively intact.

\subsection{Comparison with \sersic\ index} \label{ss:sersic}

\begin{figure}
\begin{center}
\includegraphics[width=3.4in]{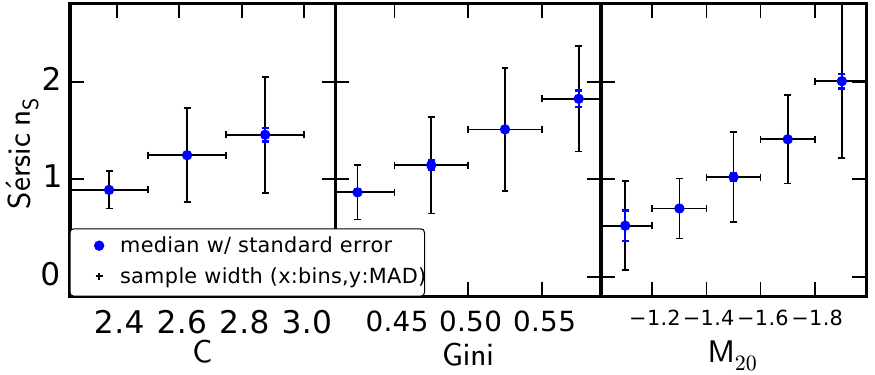}
\caption{{\sersic\ index $n_S$ versus non-parametric structural diagnostics $C$, $G$, and $M_{20}$, all measured in F160W from the same synthetic \hst\ images based on the no-RP simulations. The RP simulation morphologies are similar in this respect (Figure~\ref{fig:ssfr_and_mstar}). This shows how our simulation sample relates to other common morphology samples defined by $n_S$; there are very few, if any, analogues to low-redshift elliptical galaxy light profiles ($n_S \gtrsim 2.5$) among the \candels-detected simulation subsample.  Therefore in this paper we are primarily following the evolution of subtypes of star-forming galaxy discs.} \label{fig:sersic}}
\end{center}
\end{figure}

{ In Figure~\ref{fig:sersic} we compare measurements of \gmtwenty\ to \sersic\ index $n_S$ for a subsample of the no-RP simulations.  The relationship between \gmtwenty\ and $n_S$ has been studied in observations (e.g., M.\ Peth et al.\ in prep.), and our findings are very similar to what we would expect given previous results.  We expect that the correlation between these diagnostics is the same for the RP simulations, which have properties similar to the no-RP simulations, insofar as the galaxies emphasized here are predominately disc-dominated \citep[contrast with, e.g., compact galaxies studied by][]{Zolotov2014}. }

{ \gmtwenty\ and $n_S$ both correlate with the strength of a bulge component in a galaxy's light profile.  However, this correlation scales differently: disc-dominated galaxies pile up at $n_S \sim 1$, while bulge-dominated galaxies pile up at $G,M_{20} \sim (0.6,-2.5)$.  In other words, numerically, \gmtwenty\ has more leverage than $n_S$ to discriminate between completely disc-dominated and slightly disc-dominated, while these types of galaxies will all have $n_S \lesssim 2$.  This makes \gmtwenty\ a more appropriate morphology diagnostic for the present study of high-redshift galaxy simulations, most of which are star-forming with a substantial disc component.}

{ Figure~\ref{fig:sersic} shows that non-parametric diagnostics correlate with $n_S$ in the expected way, and that the galaxy simulation sample we study comprises mainly a diversity of galaxies which would otherwise be treated as largely disc-dominated, since they have $n_S \lesssim 2$.  Also, some bulges have exponential light profiles \citep[e.g.,][]{Carollo1999}, and so a range of different diagnostics should be applied in order to better classify galaxy structures.}

\subsection{Incompleteness}  \label{ss:depth}

For most of our mock images, the depth roughly associated with the CANDELS-wide fields was sufficient to obtain reliable morphology parameters at $z < 2$.   At higher redshift, the falling signal to noise of the mock images causes either no source detection or sources fainter than those for which we believe a reliable morphological classification is possible ($H > 24.5$).  However, some sources brighter than this magnitude cut are extremely red, causing issues owing to low signal-to-noise in our rest frame B band measurements (e.g., Z, I, V bands at $z < 1.5$).  In particular, this effect caused errors for VELA28RP (black curves ; Table~\ref{tab:sims}), which has a very red, compact bulge and very faint, large disc at $z \lesssim 1.5$.  This occurred for $\sim 2$ Gyr for this one of the ten RP simulations.  At the \candels-wide depth, the simulated source did not have enough flux in the rest-frame B filter to reliably measure $M_{20}$ and $R_P$, in particular (other quantities were well measured).  We re-analyzed its images with the background multiplied by a factor of $0.4$ (``$\sim$~1~magnitude deeper''), which increased the signal-to-noise enough to reliably measure these quantities.  In the black curves of Figure~\ref{fig:structureparams} and right-most column of Figures~\ref{fig:simimages} and \ref{fig:hstimages}, this model now reflects an extremely concentrated central bulge and extremely low surface brightness disc, almost invisible when viewed face-on.  

This brief experiment reflects the challenge in selecting and reliably classifying galaxies' shapes in a fixed rest-frame filter.  In particular, it highlights the known importance of using selection criteria independent of sources' rest-frame filter, and of applying cuts that are complete to galaxies of the desired colors and radii in addition to stellar mass.

\subsection{Resolution}  \label{ss:resolution}

\begin{figure*}
\begin{center}
\includegraphics[width=5.5in]{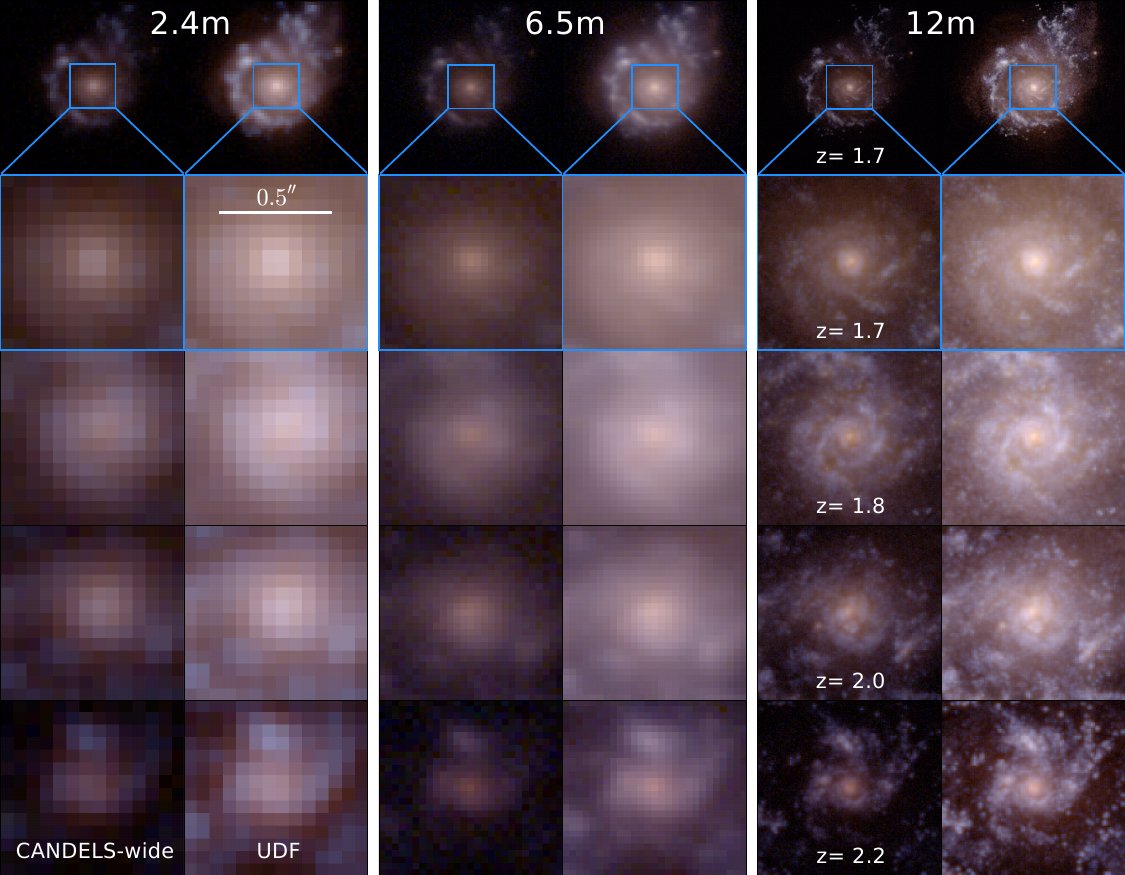}
\caption{{We demonstrate the effects of limiting surface brightness and spatial resolution on visual morphology during selected timesteps of VELA27MRP.  The top panel shows the same $\sim 5.4''$ field of view as the panels of Figure~\ref{fig:hstimages}, while the bottom panels show only the central $\sim 1''$.  Here we show colour-composite images derived from all five \hst\ filters considered in this paper. In the three column pairs, we show the spatial resolution achieved by ideal circular telescope apertures $2.4$m, $6.5$m, and $12$m, respectively.  As in Figure~\ref{fig:hstimages}, telescopes like \hst\ can capture the global morphology of this simulated galaxy at $z \sim 2$.  At higher spatial resolution, the bulges and discs break into sub-components, such as the nuclear spiral arms, star-forming knots, and dust lanes seen in the right columns.}  \label{fig:resolution}}
\end{center}
\end{figure*}

{ In Figure~\ref{fig:resolution} we show how these mock observations depend on spatial resolution at two nominal surface brightness limits.  To achieve this, we convolved the high-resolution simulated images (Figure~\ref{fig:simimages}) with a Gaussian point-spread function roughly tracking the FWHM appropriate for several ideal circular telescope apertures.  Then we added Gaussian sky shot noise to achieve $\sim 28$th and $\sim30$th magnitude point-source $5\sigma$ detection thresholds per filter, roughly matching the noise properties of CANDELS-wide and the Ultra Deep Fields, respectively. }

{ \citet{Lotz2006} and \citet{Grogin2011} simulated the structural diagnostics used throughout this section as a function of source size and brightness, and showed that they are well converged in sources as large and bright as these at \hst\ spatial resolutions and CANDELS-wide depths.  In other words, we can use the left-most column of Figure~\ref{fig:resolution} to adequately classify the broad morphological type of these simulated galaxies as measured with \gmtwenty, for example.  However, these basic structural components split into numerous sub-components when mock-observed at higher spatial resolution.  Therefore, we should be careful when using diagnostics that may be sensitive to the assumed spatial scale, including some of the merger and/or clump statistics discussed in Section~\ref{s:mergers}.}


\section{Merger Diagnostics} \label{s:mergers}

Expanding on the general structural evolution of these cosmological simulations, in this section we apply merger diagnostics to the mock images.  These qualitative studies are a precursor to a detailed study of the simulations' merger histories in combination with morphology measurements.  Some of these diagnostics may behave differently in observations (real or simulated) at higher spatial resolution (c.f., Section~\ref{ss:resolution}) and so the results of this section should at present be considered applicable only to surveys with characteristics similar to CANDELS.  

\subsection{Behavior of merger statistics in a cosmological major merger}  \label{ss:mid}

While gas accretion is thought to be the primary driver of stellar mass assembly in most galaxies in the mass range we consider ($M_*\sim10^{10} M_{\odot}$ at $z \sim 1$) galaxy interactions and mergers may be important sources of gas and may trigger significant morphological changes in some haloes.  Also, the rate of such events is observed to increase with redshift up to $z > 1$, and is predicted to be even higher at earlier times.  The simulations we study here, and others, can serve as an important test-bed for predicting how such events will be observed, and for refining diagnostics to select and study them.  Indeed, hydrodynamical simulations have enabled merger rate measurements at low redshift by predicting observability timescales of various indicators that are thought to be associated with merger events \citep{lotz08,lotz10,Lotz2010,snyder11a}.  However, such studies were limited to non-cosmological binary mergers, which most importantly do not account for the true cosmological complexity of gas accretion and merger events.

\begin{figure*}
\begin{center}
\includegraphics[width=6.0in]{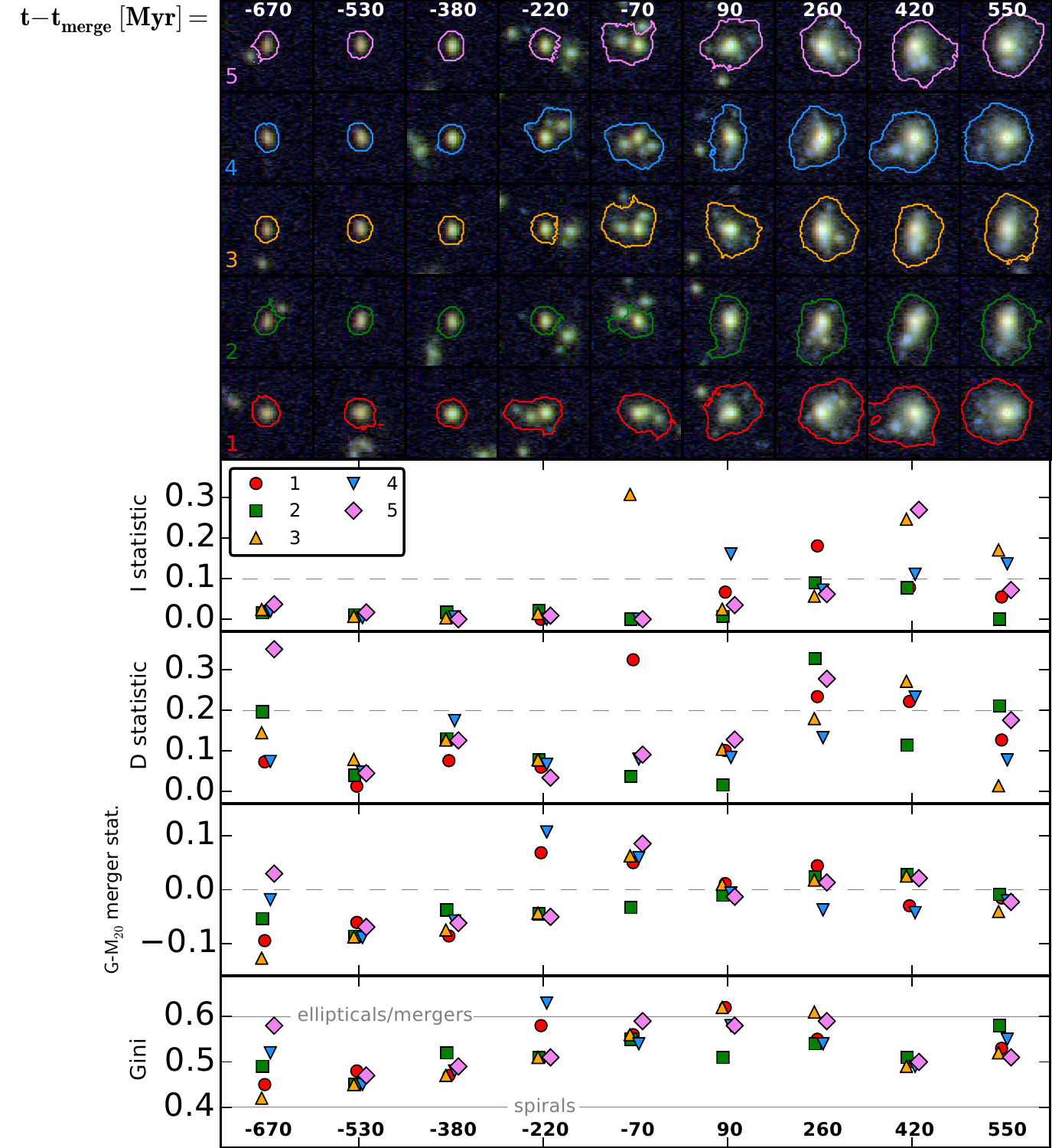}
\caption{An example galaxy merger at $z \sim 1.6$ in mock $z$-$J$-$H$ \candels\ images of simulation VELA08 (non-RP; \sunrise\ images of the RP version have not yet been calculated at the time of this writing).  This shows the same halo viewed from multiple angles over $\sim 1\rm\ Gyr$ in mock \hst\ images $\sim 3\rm\ arcsec$ across.  Colors here represent different viewing directions, not different simulations: we outline the {\sc SExtractor} segmentation maps on the images and plot the matched colored points in the graphs below.  This shows how some diagnostics are sensitive not only to the merger stage, but also to the segmentation of galaxy sources, especially at early times.  We also show a rough bulge-disc classification based on $G$-$M_{20}$: the light profile shape at $t = 550\rm\ Myr$ ($z\sim1.4$) is  largely unchanged from its initial state at $t = -670\rm\ Myr$ ($z\sim2$).    \label{fig:mergerexample}}
\end{center}
\end{figure*}

In Figure~\ref{fig:mergerexample}, we explore automated merger classifications.  The $M$, $I$, and $D$ statistics were designed to automatically identify mergers in a way that reproduces the results of visual classifications \citep{Freeman2013}.  The D statistic traces similar features as Asymmetry.  In addition, we use \gmtwenty\ to identify galaxies deviating perpendicularly from the bulge--disc locus, the evolution along which we have studied earlier in this paper.  In detail, we define the \gmtwenty\ merger statistic to be the point-line distance from the diagonal line that identifies mergers at low redshift: $G = -0.14 M_{20} + 0.33$ \citep{Lotz2004}.  Objects with negative values tend to be on the bulge-disc locus, and do not reflect obvious merger signatures.  This merger statistic tends to highlight objects with multiple cores (large $M_{20}$) or merger-induced starbursts (high $G$).  Variants of a \gmtwenty\ statistic have been used, with other statistics, to track the galaxy merger rate versus redshift \citep[e.g.,][]{Lotz2011}.   

Figure~\ref{fig:mergerexample} shows the evolution of a single halo, viewed from five different directions, that experiences one or more mergers over $1$ Gyr at $z\sim 1.6$.  We define $z = 1.6$ as the merger time $t=0$.  We show mock HST $z$-$J$-$H$ composite images at the top, and encircle the \se\ source segmentation map that fed into the morphology calculation.  This segmentation is distinct from the final segmentations used to calculate either \gmtwenty\ or $MID$, but demonstrates how any segmentation procedure depends on time and viewing angle during galaxy formation events.  This uncertainty implies that care must be taken when interpreting variations, or lack thereof, in subsequent diagnostics.  

This represents a fundamental difference between how surveys and simulations segment galaxies: group finders and merger tree codes can robustly isolate and track galaxies in three dimensions, but these identities are obfuscated when projected into multiple images.  In principle, it is possible to locate a given galaxy in the final projected image after-the-fact, but there are simpler ways to enable this analysis.  Specifically, we suggest an algorithm, which could be implemented as a feature in \sunrise, to directly create ``theoretical 2-D segmentation maps'' in the same pixel sets as the mock images themselves, by projecting the theoretically segmented galaxy ID numbers. More specifically, the light from each star particle can be tagged with the galaxy or halo ID number in which it resides.  For each pixel in the final projected image, we could find the galaxy ID number that contributes most to the total brightness at some chosen wavelength. Then a useful data product would be an image whose pixels are registered to the synthetic galaxy image, and whose pixel values are the galaxy ID number chosen in the specified manner. This would allow direct cross-matching between subhalo finding and image diagnostics, and enable joint analyses of the mock-observed sources and theoretical galaxy assembly histories.

Here, we directly perform the observational segmentation procedures as done for the \candels\ non-parametric catalogs, and do not consider the theoretical catalogs of haloes, mergers, and histories.  This limits the meaning we can assign to measurements from the mock images, but does allow us to track galaxies in time as they would be measured in surveys, and to consider the robustness and applicability of various diagnostics in Figure~\ref{fig:mergerexample}.  We defined the merger time $t=0$ as exactly $z = 1.60$.  Overall, we find that the merger classifiers we consider, \gmtwenty\ and MID, are sensitive at different stages of the merger process: the \gmtwenty\ merger statistic is activated in the early stage at $t < 0$ while MID are activated for an extended period after coalescence at $t \ge 0$.  This owes in part to the segmentation issues discussed above, but since these issues are present in survey analyses, the evolving sensitivities of the diagnostics here are meaningful.  

The long-term global structure evolves from $G \sim 0.45$ at $t = -530$ Myr to $G\sim 0.52$ at $t = 550$ Myr.  The galaxy also becomes more centrally concentrated according to $M_{20}$ and $C$ at $t > 0$.  However, the \gmtwenty\ merger statistic is enhanced more strongly during the early stage ($t\sim -200$) than the late stage ($t\sim 300$).  

Thus in this case the late-stage merger signatures are obfuscated in \gmtwenty\ space by global structural evolution: the galaxy evolves in $M_{20}$ such that the \gmtwenty\ merger statistic is lowered, leading to a false negative or marginal detections of mergers in such systems.  In other words, the threshold for diagnosing a merger with \gmtwenty\ may depend on the source's initial or evolving overall structure.  To identify these mergers robustly, a separate diagnostic or visual inspection must be applied.

The M, I, and D statistics do not have the same false negatives at late times, but they are less likely to be enhanced at early times.  In our implementation, MID and \gmtwenty\ use different segmentation algorithms, a procedure to which both diagnostics are sensitive.  For example, the red and gray points are activated at $t = -70$ owing to projections yielding images with multiple apparent nuclei.  

In Table~\ref{tab:vela08}, we summarize the observed timescales for this one merger example (VELA08, one of 22 non-RP simulations). There are other interesting such events among these simulations, which we discuss in Sections~\ref{ss:midenhance} and \ref{ss:mergerdiskgrowth}, and that we plan to analyze in greater detail in a future paper.

\begin{table*}
\caption{Total timescales, in $10^6$ yr, for VELA08 major merger event in Figure~\ref{fig:mergerexample}.  These values are the sum of time spent above the threshold levels indicated in the column headers.  Values in parenthesis measure the sum in only the timesteps at $-500 < t < 0$, where $t=0$ is defined as $z=1.60$. \label{tab:vela08}}
\begin{center}
\begin{tabular}{ccccccc}
Camera & shape/segmap color & $M > 1$ & $I > 0.1$ & $D > 0.2$ & all MID & \gmtwenty\ merger \\
\hline
5 & diamond/pink & 320 (0) & 150 (0) & 320 (0) & 150 (0) & 470 (150) \\
4 & inverted triangle/blue & 0 (0) & 450 (0) & 150 (0) & 0 (0) & 310 (310) \\
3 & triangle/orange & 470 (150) & 440 (150) & 300 (150) & 300 (150) & 630 (150) \\
2 (edge-on) & square/green & 170 (0) & 0 (0) & 300 (0) & 0 (0) & 320 (0) \\
1 (face-on) & circle/red & 0 (0) & 170 (0) & 470 (150) & 0 (0) & 640 (310) \\
\hline
\end{tabular}
\end{center}
\end{table*}

In some cases, the numerical values of the MID statistics can depend on the distance to the source. This is not apparent here owing to the slowly changing ratio between physical and angular size at $1 < z < 3$, but can manifest in studies of nearby galaxies, or comparisons between low and high redshift.  As a workaround, one can redefine $M$ using Equation~\ref{eqn:ar2} or \ref{eqn:ar3}.

\subsection{MID enhancement timescales} \label{ss:midenhance}

\begin{figure*}
\begin{center}
\includegraphics[width=6.3in]{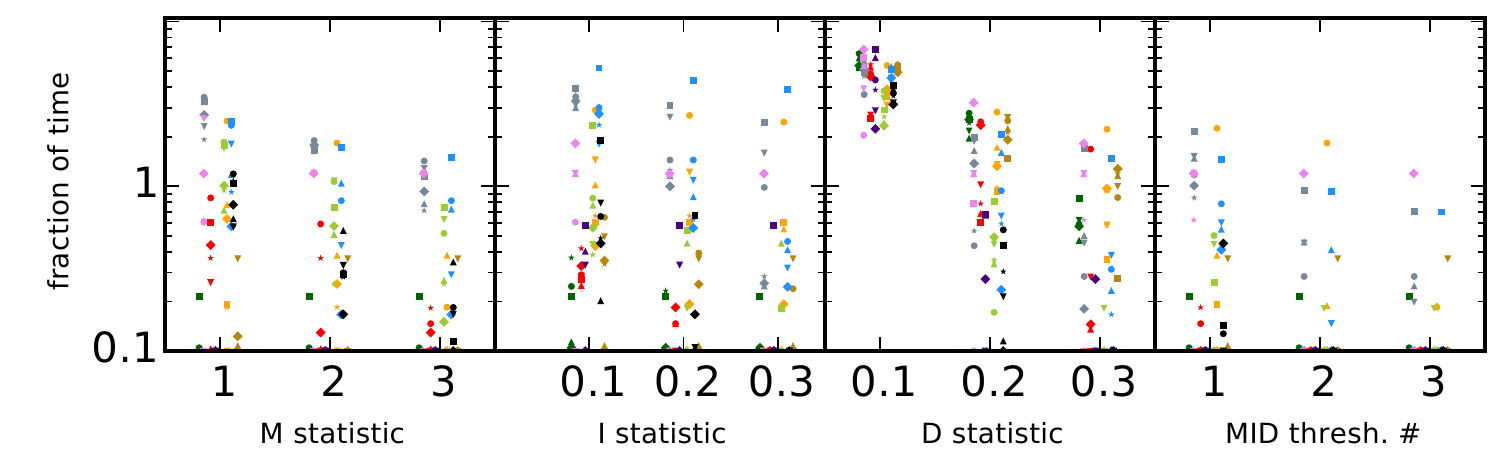}
\caption{ Fraction of time spent by the simulations above each of three thresholds in MID for the ten RP simulations studied in this paper.  The rightmost panel shows systems that are enhanced in all three of M, I, D above the 1st, 2nd, and 3rd levels of their respective panels. Colors correspond with Table~\ref{tab:sims} and to the other plots in this paper.  The six point symbols reflect the six cameras we used to ``observe'' each galaxy.  Circles are face-on, squares are edge-on; triangles, inverted triangles, and diamonds are random cameras 3, 4, and 5 as in Table~\ref{tab:vela08}; stars are another random camera.  Plotted x-axis values for a given simulation have a random shift for clarity.   The blue and gray points -- the two simulations that experience significant evolution toward disc-dominated morphologies in Figure~\ref{fig:structureparams} -- are near the top of the distribution here, suggesting that the formation of discs at $z \gtrsim 1$ is linked to higher MID statistics.  This likely owes to enhanced merger activity, the sensitivity of MID (especially the $I$ statistic) to bright clumps, or both.  \label{fig:MIDfractions}}
\end{center}
\end{figure*}

Figure~\ref{fig:MIDfractions} presents the fraction of time that the ten RP simulations spend above selected thresholds in the MID statistics.  We focus on the fewer RP simulations owing to their much higher mock image time sampling.  Thus the merger example in Figure~\ref{fig:mergerexample} is not shown, but in any case it is not a significant outlier in the sense of its duration with enhanced MID.  However, we notice two simulations which are positive outliers in this sense: the blue and gray points are at or near the maximal MID enhancement fractions in each parameter and threshold.  The gray points are simulation VELA15MRP, and blue are VELA27MRP.  Recall that these same two simulations follow a ``low Gini'' track in Figure~\ref{fig:structureparams}, and are two examples (of ten) that evolve away from the bulk trend toward more bulge-dominated and instead become more disc-dominated with time.  They are therefore also outliers in Figures \ref{fig:versus_z}, \ref{fig:mass_and_z}, and \ref{fig:ssfr_and_z}.  We discuss this connection in greater detail in Section~\ref{ss:mergerdiskgrowth}.

Other features in Figure~\ref{fig:MIDfractions} include: 1) up to roughly half of the simulations experience MID enhancement for a negligible fraction of time ($\sim 1\%$), 2) all simulated galaxies spend $\sim 50\%$ of their time with $D > 0.1$, implying that this low level of disturbance is common, 3) square points (edge-on views) tend to be MID-enhanced for longer than the circles (face-on) and other shapes, at least for the blue and gray points, suggesting that sources viewed edge-on are more likely to be identified as MID-enhanced than those viewed face-on.  

\begin{figure*}
\begin{center}
\includegraphics[width=6.3in]{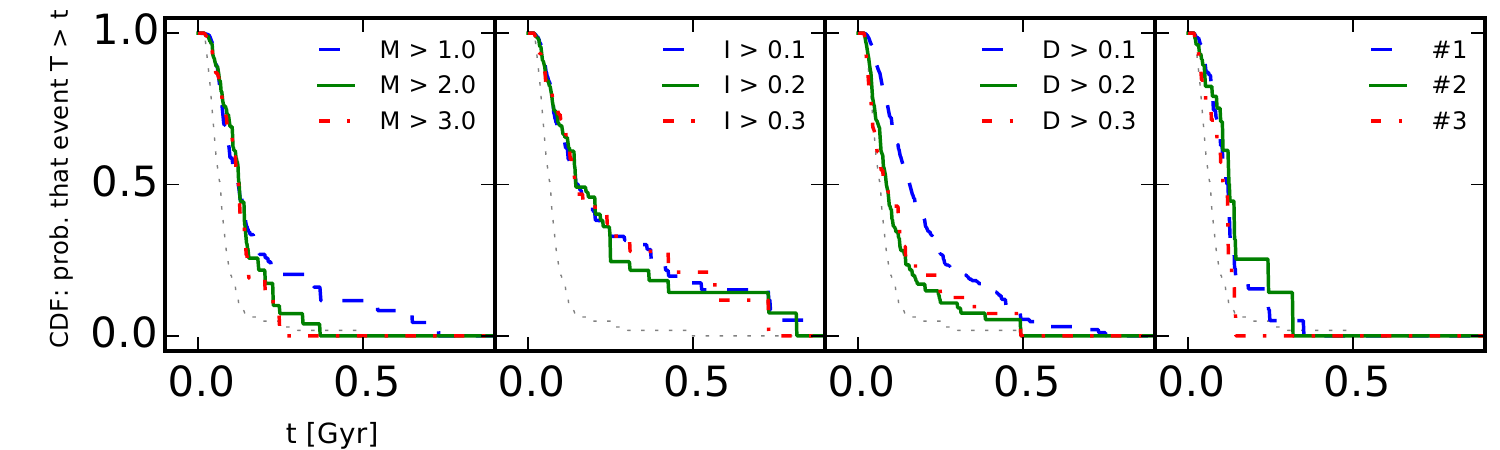}
\caption{ Cumulative distribution function (CDF) of observed event durations above various threshold levels in MID, for the ten RP simulations studied in this paper.  The dashed gray curve represents the CDF of the simulation timesteps. As an example, the blue curve in the left panel shows the probability that a galaxy observed with $M > 1.0$ will sustain at least $M = 1.0$ for $t$ Gyr during the current event. At least roughly $50\%$ of the observed events in these sets are consistent with a duration shorter than the available simulation timestep spacing.  These are a combination of complex assembly histories, potentially clumpy or disturbed morphologies that are extremely short-lived, and noise. This motivates the need for significant cosmological statistics in simulations like these in order to capture the rarer longer-lived and more robust events.  If instead these turn out to be representative of typical MID-classified mergers, then this implies that significant numbers of galaxies observed with merger-like (MID-enhanced) morphologies are not drawn-out major events but may be short-lived disturbances.  \label{fig:MIDevents}}
\end{center}
\end{figure*}

Figure~\ref{fig:MIDevents} presents this MID enhancement information in the form of cumulative distribution function (CDF) of event durations.  Here, we identify individual events during a simulated galaxy's history in which the MID statistics reside above certain thresholds, the same as in Figure~\ref{fig:MIDfractions}, as viewed from each of the six cameras.  For each threshold level, each event has a duration $t$ given by the sum of the simulation timesteps that constitute the event as viewed by a given camera.  Thus these measurements correspond to how observed sources might be expected to appear if we could observe them as they evolved in time.  When the quantity drops below the threshold level, then the event ends.  Note that this accounting separates the merger in Figure~\ref{fig:mergerexample} into at least two distinct events.  

The CDFs thus answer the following question:  ``If I observe a quantity $Q$ with $Q > Q_l$, what is the probability that $Q > Q_l$ for a duration $t$ or longer in the current event?''  The answer to this question can only be predicted with a suite of hydrodynamical simulations plus mock observations.  We show curves for three example threshold levels $Q_l$ in each quantity.  In a gray dashed curve, we plot the CDF of the time spacing between simulation output times measured in the synthetic images.  

Roughly $50\%$ of the total time experiencing enhanced MID is composed of very short ($\ll 100$ Myr) events, with CDF curves very close to the simulation time spacing CDF.  For this small set of ten simulations -- chosen to have no major mergers during the $\sim 1$ Gyr before $z = 1$ -- it appears that other events, such as minor mergers or large star-forming clumps, can enhance MID with very short durations for total duty cycles greater or comparable to longer, more obvious merger events such as the ones in Section~\ref{ss:mid} and \ref{ss:mergerdiskgrowth}.  This preliminary estimate of the observability timescales for the MID statistics implies that significant care must be taken when interpreting any small number of observations or simulations.

\subsection{Merger statistics and disc growth} \label{ss:mergerdiskgrowth}

Here we connect the two major sections of this paper.  In Section~\ref{s:morphs} we inspected the average morphology evolution of cosmologically simulated galaxies. We found that many galaxies with $9.5 < \log_{10} M_{*}/M_{\odot}< 10.5$ evolve from being disc-dominated according to \gmtwenty\ at $z \gtrsim 2$ to having a wider distribution -- more bulge-dominated galaxies -- at $z \sim 1$.  At least two simulated galaxies near the middle of this range are obvious outliers from this trend by $z \sim 1.5$ -- their rest-frame $B$-band light becomes more disc-dominated with time, with obvious star-forming discs.  Several other simulations have discs that become brighter but are less disc-dominated (higher $G$).  \citet{Zolotov2014} showed that gas discs can re-form around compact bulges that recently formed from the unstable collapse (`compaction') of a star-forming disc, and we are likely witnessing the aftereffects of such evolution and subsequent star formation through \gmtwenty. 

\begin{figure*}
\begin{center}
\includegraphics[width=6.8in]{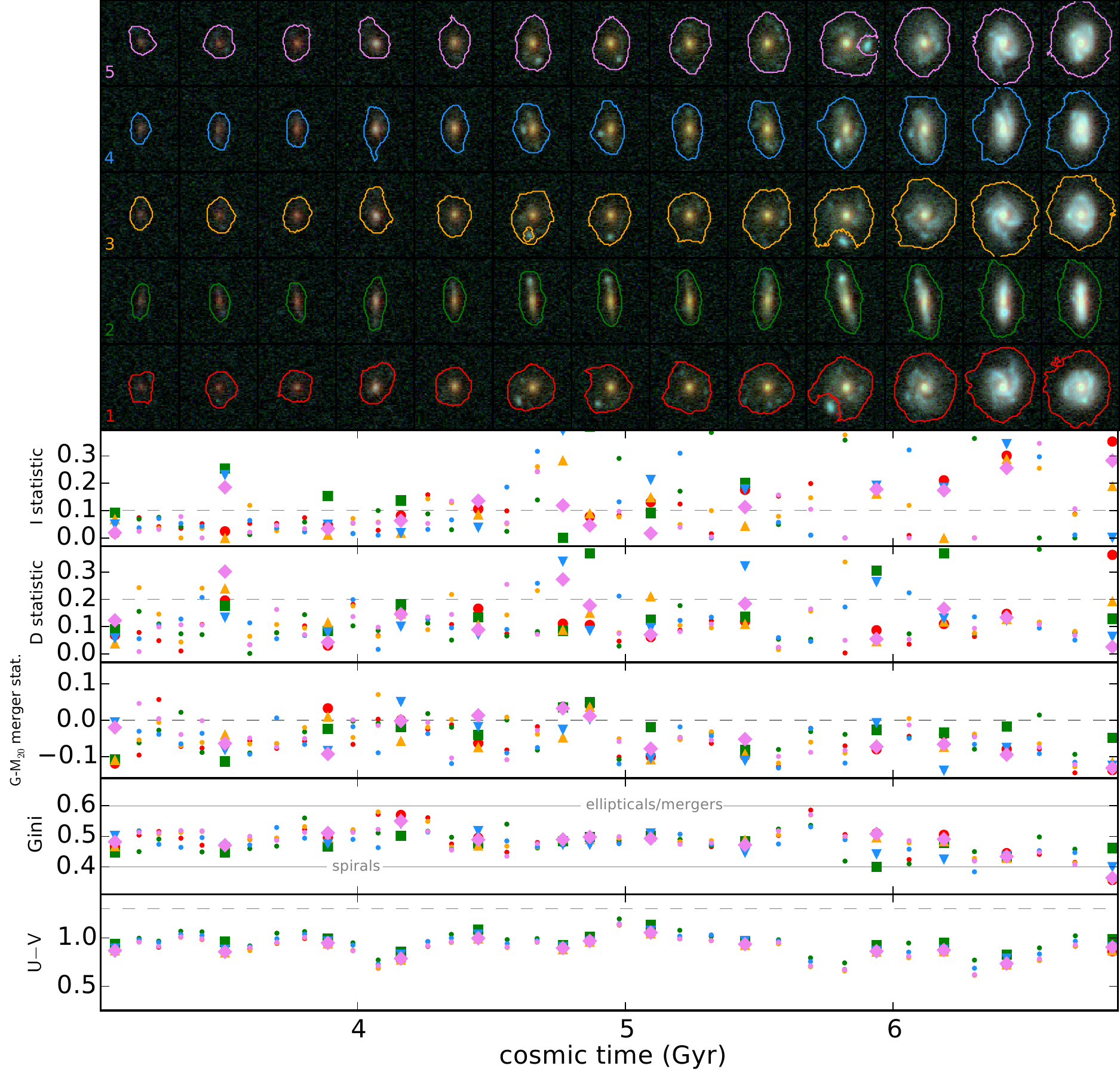}
\caption{In the spirit of Figure~\ref{fig:mergerexample}, we show the disc enhancement following a minor merger at $t \sim 6$ Gyr ($z\sim 1$) in VELA15MRP, plotted as the gray points elsewhere.  This and Figure~\ref{fig:VELA27} highlight the two simulations we identified as evolving from bulge- to disc-dominated morphologies over time (Figure~\ref{fig:structureparams}), and which spend a significant ($10\%$) of their time as MID-enhanced from Figure~\ref{fig:MIDfractions}.  The points in the bottom four graphs are as in Figure~\ref{fig:mergerexample} and Table~\ref{tab:vela08}.  At the top, from each time bin spanned by the sub-panel, we show the set of images at the simulation timestep which maximizes the sum of the $I$ statistic.  This serves to highlight the objects which contribute most to the high MID-enhanced fractions in Figure~\ref{fig:MIDfractions}.  The bright blue clump at $t \sim 5.8$ has a merger origin that is clearly established by inspecting the time evolution of the original mock images at the wider total field of view: images shown here are zoomed to show galaxy detail.  \label{fig:VELA15}}
\end{center}
\end{figure*}

\begin{figure*}
\begin{center}
\includegraphics[width=6.8in]{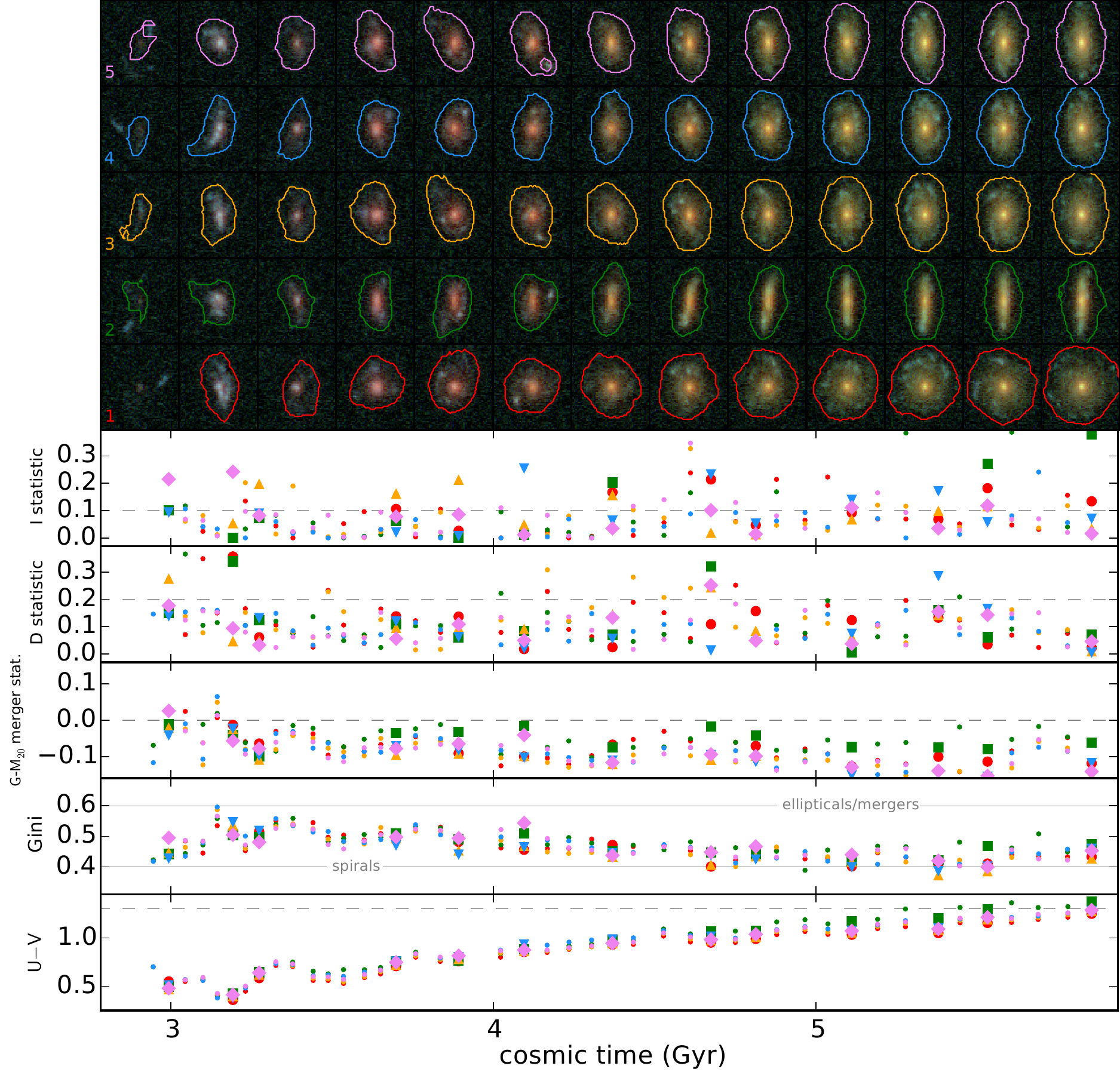}
\caption{Following Figure~\ref{fig:VELA15}, we show VELA27MRP, plotted as the blue points elsewhere.  This and Figure~\ref{fig:VELA15} highlight the two simulations we identified as evolving from bulge- to disc-dominated morphologies over time (Figure~\ref{fig:structureparams}), and which spend a significant amount ($10\%$) of their time as MID-enhanced in Figure~\ref{fig:MIDfractions}.  At $t \sim 3$ we see the late stages of a faint merger event; prior to this the source was undetected, but the merger origin of the early features is clearly established following the same visual inspection methodology as in Figure~\ref{fig:VELA15}.  Shortly thereafter at $t \sim 3.2$, we witness significant bulge growth evidenced by a red core appearing and Gini increasing from 0.4 to 0.55.  In the following $\approx 3$ Gyr, this source experiences steady disc formation and evolution.  At $t \sim 4.5$, this source experiences MID enhancement, possibly owing to its clumpy nature.  \label{fig:VELA27}}
\end{center}
\end{figure*}

In Section~\ref{ss:midenhance} we noticed that these same two simulations spend the highest fraction of their time with enhanced MID statistics, especially viewed from the edge-on direction.  We explore these two examples in greater detail in Figures~\ref{fig:VELA15} (VELA15MRP; gray symbols elsewhere) and \ref{fig:VELA27} (VELA27MRP; blue symbols elsewhere).  These figures follow Figure~\ref{fig:mergerexample} by showing at the top composite color images of the same simulation timesteps from five camera angles, where the bottom two are the face-on and edge-on orientations (red and orange segmentation maps, respectively).  We zoom in to the sources such that the field of view of each panel is $3.6$ arcsec across.  At the bottom we show several morphology statistics: $I$, $D$, \gmtwenty\ merger statistic (Section~\ref{ss:mid}, and $G$.  We do not show the $M$ statistic because it behaved similarly to $I$ and $D$ in Figure~\ref{fig:mergerexample}.  These RP simulations have very fine time sampling, and so we show all points for the morphology measurements, but sub-sample the images we show at the top of the figure.  We do this by dividing the full time range into bins spanned by the individual image panels.  From each bin, we select the timestep with the maximal sum of the $I$ statistic over all six camera angles, display those images in the top section, and show a larger plot symbol for the selected timestep in the bottom section.  This allows us to display all quantitative morphology measurements while enabling us to compactly visualize the timesteps with enhanced $I$.  

First, since many of the simulations are MID-enhanced for longer when viewed-edge on, large clumps in the discs may contribute to these enhancements, and their origin may be in-situ or ex-situ \citep[][Mandelker et al.\ in prep.]{Mandelker2014}.  This could arise from the increased line-of-sight star formation intercepted by these cameras observing clumpy star-forming discs \citep{Moody2014}.  Indeed, several of the I-enhanced images of Figures~\ref{fig:VELA15} and \ref{fig:VELA27} (and some timesteps not shown) appear to be selecting in-situ clumps.  This is true at $t \sim 4.8$ Gyr and $t \sim 6.5$ Gyr in Figure~\ref{fig:VELA15}, and $t \sim 4.5$ Gyr and $t \sim 5.5$ Gyr in Figure~\ref{fig:VELA27}.  Thus we conclude that MID may in some cases be selecting clumpy, growing discs from \hst\ images at $1 \lesssim z \lesssim 2$.  { The link between the MID statistics and observationally identified giant clumps \citep[e.g.,][]{Guo2015} is unclear, in part because we have focused on measuring the MID statistics at slightly different rest-frame wavelengths (optical) than for which \citet{Guo2015} identifies giant star-forming clumps (UV). }  

At other times, MID is clearly triggered by mergers, such as at $t = 5.95$ Gyr in Figure~\ref{fig:VELA15} and $t \sim 3$ Gyr in Figure~\ref{fig:VELA27}.  In the former, the VELA15MRP simulation shows a bright blue region for a very short period ($< 100$ Myr).  By inspecting the time evolution of individual mock images at their full field of view ($\sim 10$ arcsec or $\sim 80$ kpc) as opposed to the zoomed versions shown in this paper, we robustly identified this bright blue region as a small galaxy that is merging with the primary galaxy.  This and nearby timesteps have enhanced $G$, $I$, and $D$ statistics (also $M$, $C$, and $A$, not shown), while the \gmtwenty\ merger statistic increases but remains below $0$ through this period.  This lack of a \gmtwenty\ merger signature reflects the generally low $G$ value of this disc-dominated source: as in Section~\ref{ss:mid}, we see that the ideal \gmtwenty\ merger classification threshold may depend on the underlying morphology of the system.  

This minor merger occurs at an interesting time during the total evolution of VELA15MRP: at $t \lesssim 5$ Gyr, this galaxy has a compact, red core with a faint disc.  The galaxy is still forming stars, and is never ``quenched'' in terms of a very low SSFR ($\lesssim 10^{-11}\rm\ yr^{-1}$).  This early-forming core is stable, with roughly the same brightness at $ 4 < t < 7$ Gyr.  The faint star-forming disc appears stable until $t = 6$ Gyr, just after the merger event, after which it experiences a disc-wide starburst event.  This event then dominates the total color and brightness of the source by $t \sim 7$ Gyr ($z \sim 1$): the compact core remains at seemingly the same brightness and size, while the disc grows vigorously around it.  Contrast this with the steadiness of the discs in other RP simulations at $z \sim 1$: Figures~\ref{fig:VELA27} and \ref{fig:hstimages}.  While we cannot conclusively identify the merger as the causal trigger of significant disc star formation here, it is an intriguing suspect.  It is possible that it disrupts or adds to the galaxy's gas reservoir in such a way as to enable significant star formation in the disc, or that the smooth accretion of gas is enhanced during this event.  

Figure~\ref{fig:VELA27} shows VELA27MRP, the other RP simulation with a disc-dominated light profile that becomes more disc-dominated with time.  Here, the structure of the source is roughly stable at $t \gtrsim 4$ Gyr (though the disc may be growing in size here).  However, prior to this, this halo experiences a major merger at $t \sim 3$ Gyr ($z \sim 2.2$), as seen not only by the first two disturbed image panels but also by inspection of the time evolution of wider-field images prior to these: we robustly identify a merger event here.  Indeed, before $t \sim 2.8$ Gyr this source is undetected in our \candels-wide-like photometry, and before $t \sim 3$ it has a magnitude $H > 24.5$ that disqualifies it from our morphological pipeline.  Its H magnitude increases by at least $\sim 2$ mags from $z=2.4$ to $z=2.1$ and fluctuates by $\sim 0.5$ mag shortly thereafter, reflecting evolution after a major starburst event ($\rm SFR \sim 100 M_{\odot}\ yr^{-1}$).  This is supported also by the blue points in Figure~\ref{fig:globalparams} that are well above the SF main sequence at $z \sim 2.2$.  

After the merger, at $t \sim 3.5$ Gyr, this source obtains a red bulge with properties -- like VELA15MRP above -- roughly constant with time for several Gyr, surrounded by a faint star-forming disc.  This is seen not only in the images but also the values of $G$: $G > 0.5$ at $t \sim 3.5$ Gyr shrinking to $G \sim 0.4$ at $t \sim 6$ Gyr as the disc begins to dominate at later times.  In this case it is even harder to identify the merger as the cause of disc formation.  However, it is possible that it is the cause of bulge growth at early times and may have played a role in allowing the disc to grow at later times, either by triggering the subsequent SF, delivering more gas, or both.  The simulation we identified as having a large undetected disc (VELA28MRP; black symbols elsewhere; Section~\ref{ss:depth}) evolves in a manner very similar to this example (VELA27MRP), but for brevity we do not show its evolution.  

{Overall, timesteps with visually obvious mergers or extremely clumpy star formation are relatively rare for the subset of simulations studied here (Figure~\ref{fig:simimages}).  During the stable periods, simulated galaxies tend to have spiral and bulge structures similar to those common in the low-redshift universe. } 

\subsection{Morphology versus Color}  \label{ss:color}
\begin{figure*}
\begin{center}
\includegraphics[width=5.5in]{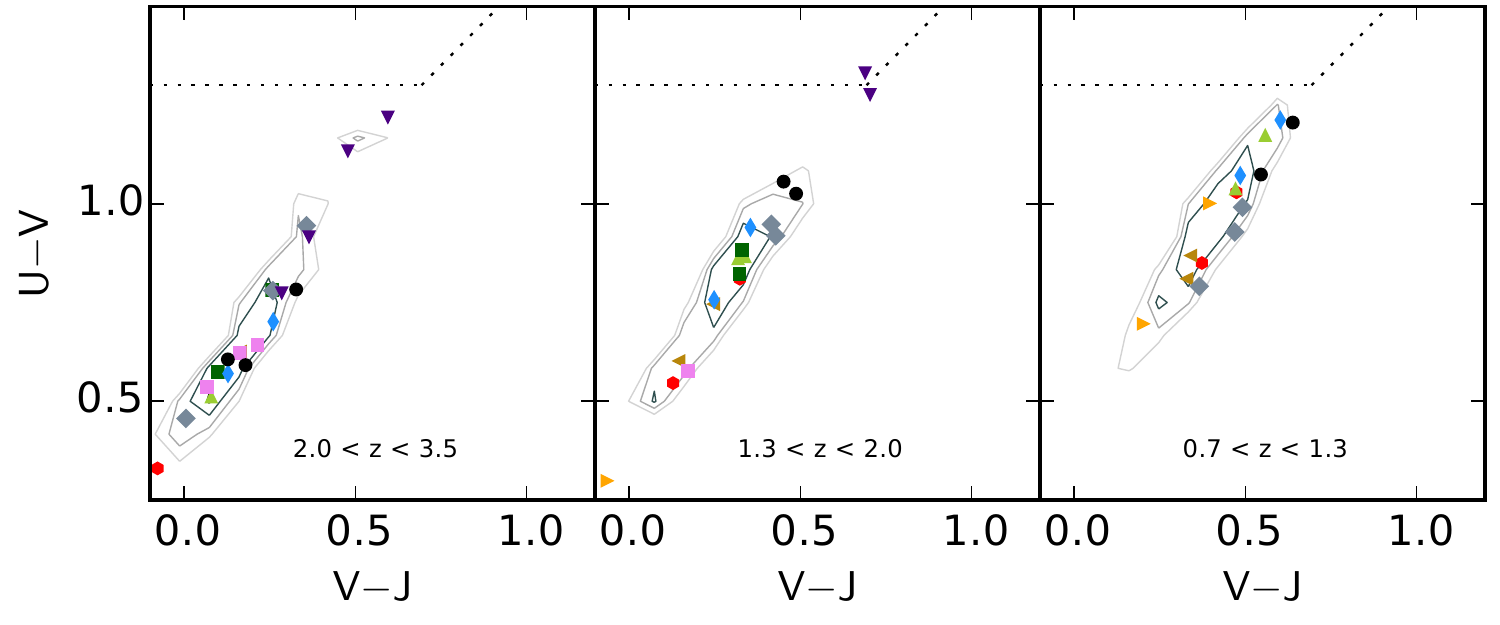}
\caption{{Following Figure~\ref{fig:versus_z}, we show rest-frame $UVJ$ colors, plotted in magnitudes, versus redshift for the ten RP simulations, with dotted lines a nominal selection for passive galaxies \citep[e.g.,][]{Williams2009,Patel2012,Whitaker2013}.  This highlights the fact that we study here primarily star-forming simulated galaxies with average or low dust content ($V-J < 1$). As the simulations evolve to earlier-type values of \gmtwenty\ (Figure~\ref{fig:versus_z}), they also gradually become redder, on average.  }  \label{fig:uvj_versus_z}}
\end{center}
\end{figure*}

{In Figure~\ref{fig:versus_z}, we showed the general trend that these simulated galaxies evolve with time toward earlier-type (less disc-dominated) optical morphologies.  At the same time, in Figure~\ref{fig:uvj_versus_z}, these galaxies evolve gradually toward the ``passive'' region of $U$-$V$-$J$ color space \citep{Williams2009}.  This picture follows the well studied and tight link between bulges and quenching \citep[e.g.,][]{Kauffmann2003,Wuyts2011,Bell2012}.  In this subset of galaxies, the $V-J$ colors are bluer than the observed galaxy population, reflecting a limited impact of dust attenuation on the stellar populations seen in these filters \citep[also seen by][]{Moody2014}. }

{However, the trajectories of individual galaxies in Figures~\ref{fig:VELA15} and \ref{fig:VELA27} tell a slightly more complex story.  At certain times, optical morphology tracks color in this same sense: as the morphology becomes more bulge-dominated, the color becomes redder, and vice-versa.  For example, this appears to be true at $t\sim 3.7$ and $t\sim 6.2$ Gyr in Figure~\ref{fig:VELA15}, and at $t\sim 3.4$ Gyr in Figure~\ref{fig:VELA27}.  However, there are other times where this connection inverts: over several Gyr, VELA27MRP (Figure~\ref{fig:VELA27}) becomes redder yet more disc-dominated with time, the opposite of the expected trend.  There are periods for which $G$ and $U-V$ evolve in opposite directions for short periods of time in VELA15MRP (VELA15MRP), for example $t\sim 4$ Gyr and $t \sim 5.7$ Gyr.}

{Thus, as a population, these simulated galaxies evolve such that their colors become redder and their morphologies become less disc-dominated, but the behavior of individual galaxy trajectories around this average trend can vary wildly.}

\section{Implications} \label{s:discussion}

\subsection{Connection between morphology, mass, and star formation}
In the stellar mass range $10^{9.3} \lesssim M_* / M_{\odot} \lesssim 10^{10.7}$, we find that galaxies in the haloes selected here experience a wide range of structural evolution.  While the majority of the simulated galaxies in this range tend to become more bulge-dominated with time, with increasing $C$ and $G$, and decreasing $M_{20}$, over $20\%$ of our simulations evolve in the opposite sense. Recall from Section~\ref{ss:hydroart} that these were selected to have no major mergers ongoing at $z = 1$ (up to roughly $\sim 1$ Gyr prior), so this finding applies to only a subset of galaxies. However, since it is the subset with possibly fewer merger events, then we expect that the diversity of morphology tracks experienced in these galaxies at $1 < z < 2$ should be considered a lower limit to the variation experienced by the true galaxy population. { At least one case (Figure~\ref{fig:VELA15}) of rapid disc brightening appears to be caused by a minor merger event, and this is reminiscent of theoretical models in which a prominent disc component can grow or re-grow owing to a gas-rich major merger or accretion event \citep[e.g.,][]{robertson06,Governato2009,Brennan2015}.  }

We find that some (at least 2/5) simulated galaxies in the top half of this mass range ($M_* \gtrsim 10^{10} M_{\odot}$) become more disc-dominated over time at $1 < z < 2$ (Figures~\ref{fig:hstimages} and \ref{fig:structureparams}).  While not unexpected, this implies many galaxies at higher redshift ($z \sim 3$) may not evolve monotonically toward progressively earlier types at lower redshift. This must be considered when analyzing the demographics of quenching over cosmic time: galaxies observed to be quenched and/or compact may not always remain so.  Indeed, \citet{VanderWel2014} find evidence for disc formation in massive galaxies at high redshift.  Of course, at these or somewhat higher masses, other forms of feedback -- such as from SMBHs -- may become important \citep[e.g.,][]{croton06} and may prevent the growth of a star-forming disc in these cases.

Broadly speaking, the correlations of mass and SFR with observed morphology (measured by \gmtwenty) are the same in both the no-RP and RP feedback cases (Figures~\ref{fig:mass_and_z} and \ref{fig:ssfr_and_mstar}; Section~\ref{ss:rptest}).  This implies that the most important factors in determing the existence of these correlations are realistic assembly histories and some self-regulation of star formation \citep[e.g.,][]{Agertz2011, Hopkins2013a, Stinson2012, Marinacci2013}.  The details of the model implementation will not change significantly how morphology measured with \gmtwenty\ at $\sim$ kpc-scale resolution will trace the galaxy's mass and SFR.  Specific model choices can do better (or worse) at matching the distributions of mass and SFR \citep[e.g.,][]{Trujillo-Gomez2013}, in which case we predict that the number densities of galaxies as a function of \gmtwenty\ will also be better (or worse) match.  Also other types of measurements, such as detailed shapes \citep{VanderWel2014} and/or clump statistics \citep{Moody2014}, likely depend strongly on the specific feedback implementation (C14).

With powerful feedback resulting from massive stars and supernovae, these models retain significant gas reservoirs. At low masses, the strong radiative feedback of C14 inhibits star formation at late times \citep{Trujillo-Gomez2013}, leading to fewer low-mass models with $G \lesssim 0.45$ (very disc-dominated) in the RP simulations compared with the no-RP simulations (at least at $z > 1.8$; Figure~\ref{fig:mass_and_z}).  The continued disc growth among the more massive galaxies in both sets owes to the fact that supernova or strong radiative feedback cannot completely prevent star formation in such haloes.  This gives rise to increasing star formation efficiency as a function of halo mass, peaking at $M_* \sim 10^{10}$--$10^{11} M_{\odot}$, $M_{halo} \sim 10^{12} M_{\odot}$ \citep[e.g.,][]{Mandelbaum2006,Conroy2009,Moster2010,Behroozi2013}.  The details of this trend may be imperfectly predicted by the assumed feedback models, e.g., they may lack some impact of AGN feedback or other quenching mechanisms.

\subsection{Mergers}

The response of these galaxies and their haloes to mergers appears difficult to generalize.  We identified several merger events and find that while they are associated with bulge growth, they are also associated with disc growth in the subsequent $\sim 1-3$ Gyr.  Image diagnostics such as \gmtwenty\ and MID respond in various ways depending on the orbit, mass ratio, and viewing angle of the megers, and they appear very sensitive to the segmentation algorithm, which during complex merger events can scramble galaxy identities.  

The newly proposed MID statistics \citep{Freeman2013} are activated for an extended period ($\sim 2-5 \times 10^8$ yr) during an obvious merger.  However, the majority of MID enhancement events in these simulated galaxies are $< 100$ Myr, approaching the measurement limit set by the stored simulation time spacing.  

MID are enhanced during the periods of rapid disc formation.  In some cases, these enhancements reflect minor merger events (potentially contributing to the disc formation), but in others they identify in-situ clumps of star formation. Some of these clumpy events are associated with mergers, while others appear to occur in isolation or long after a merger or accretion event (Y.\ Guo et al.\ in prep.).  Since episodes of significant star formation are common in such galaxies, observed distant galaxies often have patchy or clumpy morphologies \citep[e.g.,][]{Guo2012b,Guo2015}.  The present simulations experience similar phenomena \citep[e.g.,][]{Moody2014,Mandelker2014}.  At present, it is unclear the extent to which otherwise isolated clumpy galaxies might actually be associated with minor mergers.  An essential step toward disentangling these processes is to identify and calibrate a diverse suite of diagnostics against realistic simulations.  Our goal for subsequent work is to cross-match the detailed histories of these haloes with the suite of image diagnostics.  This will enable us to derive ideal diagnostics and/or combinations of diagnostics to identify high-redshift galaxy mergers.

Moreover, comparison between the detailed merger histories of these simulations and synthetic images will enable us to determine how best to incorporate treatments of merger observability into simpler models of galaxy formation.  For example, by linking image diagnostics to merger trees, we can predict the distribution of observed merger outcomes in semi-analytic or analytic models, allowing us to probe a broader set of assembly histories than are available in hydrodynamical simulations.

\subsection{Future mock image science}

The conclusions above imply that a large volume of simulated cosmological histories may be required to span the range of these outcomes and thereby accurately calibrate these diagnostics for use with high-resolution imaging of objects at $z > 1$.  It is not known if or by how much a realistic cosmological sampling of galaxy interactions may alter the observability timescales of merger induced morphological disturbances or starbursts \citep[e.g.,][]{lotz08,snyder11a}.  And we have not yet performed a full accounting of the origin of the galaxy structures we observe forming in the present simulations: i.e., how observed structures relate to the full accretion and interaction histories of these haloes.  Moreover, it is not yet clear how many simulated galaxies we require for this endeavor, but it is likely that we need at least several examples each spanning the desired range of density and mass.  Thus we are motivated to analyze not only more of the high-resolution simulations discussed here, but also much larger mock image datasets that traces the morphology of each of thousands of galaxies at a variety of masses \citet{Torrey2015}, even if their morphologies and time evolution are more coarsely resolved.  Indeed, recent studies find that the combination of supernova and AGN feedback may naturally set the morphology--mass--star formation correlations in galaxy populations at low redshift (Snyder et al.\ in prep.).  As the methods and availability of computing resources increase, we will be able to exploit not only statistical samples of models, but also finely detailed modeling in time and space.  

\section{Conclusions} \label{s:conclusion}

We studied the emergence of basic morphological types in galaxies simulated in a cosmological context.  Applying dust radiative transfer and stellar population synthesis to high-resolution cosmological hydrodynamical simulations, we observed the simulated galaxies as if they were sources detected by the \candels\ survey with \hst.  We then characterized their rest-frame optical structures as a function of time, mass, and star formation rate.  This represents one of the most direct applications of cosmological simulations toward the interpretion of state-of-the-art observations of galaxy structure across cosmic time, setting the stage for statistical comparisons between galaxy formation surveys and simulations using identical tools.

In the simulations studied here, having ${9.3} < \log_{10} M_*/M_{\odot} < {10.7}$ at $z\gtrsim 1$, we find:
\begin{enumerate}
\item{On average, these simulated galaxies show more bulge-like light profiles as time passes. Quantitatively, the average locus of at least half of the model galaxies evolves from $(G,M_{20})\sim (0.47,-1.7)$ at $z\gtrsim 2$, overlapping with late type galaxies, to $\sim (0.55, -2.0)$ at $z \sim 1$, entering the space occupied by observed early-type galaxies (e.g., E, S0, Sa), while not yet reaching the region occupied by $z\sim 0$ massive ellipticals $\sim (0.6, -2.5)$.   }
\item{Galaxies simulated with and without this implementation of radiation pressure (RP) feedback both appear to have similar correlations of stellar mass and SFR with \gmtwenty. Therefore, the relationship between current star formation and some measures of galaxy structure in \hst\ images of distant galaxies can be somewhat insensitive to the details of the feedback model.  However, this model's feedback does not include IR trapping, and therefore stronger models may lead to a larger effect. }
\item{Several of the more massive galaxies ($M_* > 10^{10} M_{\odot}$) have rest-frame optical light profiles that become more disc-dominated with time, implying that galaxies with bulge-dominated light profiles may not remain so forever.  }
\item{A wide distribution of \gmtwenty\ morphologies is apparent at $M_* > 10^{10} M_{\odot}$ and $SSFR < 10^{-9.7}\rm\ yr^{-1}$: the morphology distribution appears roughly bimodal between earlier and later types as defined by \gmtwenty\ in the rest-frame $B$ filter.  Normal star-forming galaxies ($SSFR \sim 10^{-9}\rm\ yr^{-1}$) have predominantely disc-dominated light profiles.  }
\item{Merger diagnostics of cosmologically simulated galaxies are very sensitive to the merger stage and to the segmentation algorithm, or how nearby objects are identified and separated. }  
\item{The observability timescales of a generic event (either mergers or clumpy star formation) in the newly proposed M, I, and D statistics are on average very short ($< 100$ Myr), at least among this initial sample of assembly histories.  }
\item{In a major merger simulated at $z\sim 1.6$ and measured with default segmentation algorithms, the newly proposed M, I, and D statistics are activated at times distinct and complementary to classical indicators based on \ginimtwenty: MID traces $\sim 0.5$ Gyr at later stages, much longer than the average event duration of $\sim 100$ Myr, while \gmtwenty\ traces $\sim 0.3$ Gyr at earlier stages.}
\item{The $\sim 2$ of 10 simulations with obvious disc formation at $1 < z < 2$ also spend roughly $10\%$ of their time with enhanced MID, reflecting giant star-forming clumps and minor mergers that contribute to disc assembly.}

\end{enumerate}  


\section*{Acknowledgements}

We thank Paul Torrey, Miguel Rocha, and Harry Ferguson for fruitful discussions that contributed to this paper, and Patrik Jonsson for writing, developing, and supporting the \sunrise\ code.  We thank Priya Kollipara and Yicheng Guo for contributing to the mock observation analysis pipeline and database, and the anonymous referee for helpful suggestions.

The simulations were performed at NASA Advanced Supercomputing (NAS) at NASA Ames Research Center and at the National Energy Research Scientific Computing Center (NERSC), Lawrence Berkeley National Laboratory. This work was partially supported by MINECO-AYA2012-31101. DC  is a Juan de la Cierva fellow. This research has been partly supported by ISF grant 24/12, by NSF grant AST-1010033, and by the I-CORE Program of the PBC and the ISF grant 1829/12.  GS and JL appreciate support from the \hst\ grants program, number HST-AR-$12856.01$-A. Support for program \#12856 (PI J.\ Lotz) was provided by NASA through a grant from the Space Telescope Science Institute, which is operated by the Association of Universities for Research in Astronomy, Inc., under NASA contract NAS 5-26555. This work is based on observations taken by the \candels\ Multi-Cycle Treasury Program with the NASA/ESA \hst, operated by AURA under NASA contract NAS 5-26555.

This research made use of NASA's Astrophysics Data System.  The figures in this paper were constructed with the Matplotlib Python module \citep{Hunter:2007}.

\bibliographystyle{mn2e}
\bibliography{$HOME/Dropbox/library}

\begin{thebibliography}{101}
\expandafter\ifx\csname natexlab\endcsname\relax\def\natexlab#1{#1}\fi

\bibitem[{Abraham {et~al}\mbox{.}(1996)Abraham, Tanvir, Santiago, Ellis,
  Glazebrook, \& van~den Bergh}]{Abraham1996}
Abraham R.~G., Tanvir N.~R., Santiago B.~X., Ellis R.~S., Glazebrook K.,
  van~den Bergh S., 1996, \mnras, 279

\bibitem[{Abraham, van~den Bergh \& Nair(2003)Abraham, van~den Bergh, \&
  Nair}]{Abraham2003}
Abraham R.~G., van~den Bergh S., Nair P., 2003, \apj, 588, 218

\bibitem[{Agertz, Teyssier \& Moore(2011)Agertz, Teyssier, \&
  Moore}]{Agertz2011}
Agertz O., Teyssier R., Moore B., 2011, \mnras, 410, 1391

\bibitem[{Behroozi, Wechsler \& Conroy(2013)Behroozi, Wechsler, \&
  Conroy}]{Behroozi2013}
Behroozi P.~S., Wechsler R.~H., Conroy C., 2013, \apj, 770, 57

\bibitem[{Bell {et~al}\mbox{.}(2012)Bell, van~der Wel, Papovich, Kocevski,
  Lotz, McIntosh, Kartaltepe, Faber, Ferguson, Koekemoer, Grogin, Wuyts,
  Cheung, Conselice, Dekel, Dunlop, Giavalisco, Herrington, Koo, McGrath,
  de~Mello, Rix, Robaina, \& Williams}]{Bell2012}
Bell E.~F. {et~al.}, 2012, \apj, 753, 167

\bibitem[{Bershady, Jangren \& Conselice(2000)Bershady, Jangren, \&
  Conselice}]{Bershady2000}
Bershady M.~A., Jangren A., Conselice C.~J., 2000, \aj, 119, 2645

\bibitem[{Bertin \& Arnouts(1996)}]{bertin96}
Bertin E., Arnouts S., 1996, \aaps, 117, 393

\bibitem[{Bournaud {et~al}\mbox{.}(2011)Bournaud, Dekel, Teyssier, Cacciato,
  Daddi, Juneau, \& Shankar}]{Bournaud2011}
Bournaud F., Dekel A., Teyssier R., Cacciato M., Daddi E., Juneau S., Shankar
  F., 2011, \apj, 741, L33

\bibitem[{Brammer {et~al}\mbox{.}(2009)Brammer, Whitaker, van Dokkum,
  Marchesini, Labb\'{e}, Franx, Kriek, Quadri, Illingworth, Lee, Muzzin, \&
  Rudnick}]{brammer09}
Brammer G. {et~al.}, 2009, \apjl, 706, L173

\bibitem[{Brennan {et~al}\mbox{.}(2015)Brennan, Pandya, Somerville, Barro,
  Taylor, Wuyts, Bell, Dekel, Ferguson, McIntosh, Papovich, \&
  Primack}]{Brennan2015}
Brennan R. {et~al.}, 2015, eprint arXiv:1501.06840

\bibitem[{Bush {et~al}\mbox{.}(2010)Bush, Cox, Hayward, Thilker, Hernquist, \&
  Besla}]{bush10}
Bush S., Cox T., Hayward C., Thilker D., Hernquist L., Besla G., 2010, \apj,
  713, 780

\bibitem[{Cacciato, Dekel \& Genel(2012)Cacciato, Dekel, \&
  Genel}]{Cacciato2012}
Cacciato M., Dekel A., Genel S., 2012, \mnras, 421, no

\bibitem[{Cardelli, Clayton \& Mathis(1989)Cardelli, Clayton, \&
  Mathis}]{Cardelli1989}
Cardelli J.~A., Clayton G.~C., Mathis J.~S., 1989, \apj, 345, 245

\bibitem[{Carollo(1999)}]{Carollo1999}
Carollo C.~M., 1999, \apj, 523, 566

\bibitem[{Ceverino, Dekel \& Bournaud(2010)Ceverino, Dekel, \&
  Bournaud}]{Ceverino2010}
Ceverino D., Dekel A., Bournaud F., 2010, \mnras, 404, 2151

\bibitem[{Ceverino {et~al}\mbox{.}(2012)Ceverino, Dekel, Mandelker, Bournaud,
  Burkert, Genzel, \& Primack}]{Ceverino2012}
Ceverino D., Dekel A., Mandelker N., Bournaud F., Burkert A., Genzel R.,
  Primack J., 2012, \mnras, 420, 3490

\bibitem[{Ceverino {et~al}\mbox{.}(2015)Ceverino, Dekel, Tweed, \&
  Primack}]{Ceverino2015}
Ceverino D., Dekel A., Tweed D., Primack J., 2015, \mnras, 447, 3291

\bibitem[{Ceverino \& Klypin(2009)}]{Ceverino2009}
Ceverino D., Klypin A., 2009, \apj, 695, 292

\bibitem[{Ceverino {et~al}\mbox{.}(2014)Ceverino, Klypin, Klimek,
  Trujillo-Gomez, Churchill, Primack, \& Dekel}]{Ceverino2014}
Ceverino D., Klypin A., Klimek E.~S., Trujillo-Gomez S., Churchill C.~W.,
  Primack J., Dekel A., 2014, \mnras, 442, 1545

\bibitem[{Conroy \& Wechsler(2009)}]{Conroy2009}
Conroy C., Wechsler R.~H., 2009, \apj, 696, 620

\bibitem[{Conselice(2003)}]{Conselice2003a}
Conselice C.~J., 2003, \apjs, 147, 1

\bibitem[{Conselice {et~al}\mbox{.}(2003)Conselice, Bershady, Dickinson, \&
  Papovich}]{Conselice2003}
Conselice C.~J., Bershady M.~A., Dickinson M., Papovich C., 2003, \aj, 126,
  1183

\bibitem[{Croton {et~al}\mbox{.}(2006)Croton, Springel, White, {De Lucia},
  Frenk, Gao, Jenkins, Kauffmann, Navarro, \& Yoshida}]{croton06}
Croton D. {et~al.}, 2006, \mnras, 365, 11

\bibitem[{Danovich {et~al}\mbox{.}(2012)Danovich, Dekel, Hahn, \&
  Teyssier}]{Danovich2012}
Danovich M., Dekel A., Hahn O., Teyssier R., 2012, \apj, 422, 1732

\bibitem[{Dekel \& Burkert(2013)}]{Dekel2013a}
Dekel A., Burkert A., 2013, \mnras, 438, 1870

\bibitem[{Dekel \& Krumholz(2013)}]{Dekel2013b}
Dekel A., Krumholz M.~R., 2013, \mnras, 432, 455

\bibitem[{Dekel, Sari \& Ceverino(2009)Dekel, Sari, \& Ceverino}]{Dekel2009}
Dekel A., Sari R., Ceverino D., 2009, \apj, 703, 785

\bibitem[{Dekel {et~al}\mbox{.}(2013)Dekel, Zolotov, Tweed, Cacciato, Ceverino,
  \& Primack}]{Dekel2013}
Dekel A., Zolotov A., Tweed D., Cacciato M., Ceverino D., Primack J.~R., 2013,
  \mnras, 435, 999

\bibitem[{Draine \& Li(2007)}]{draine07}
Draine B., Li A., 2007, \apj, 657, 810

\bibitem[{Dwek(1998)}]{dwek98}
Dwek E., 1998, \apj, 501, 643

\bibitem[{Freeman {et~al}\mbox{.}(2013)Freeman, Izbicki, Lee, Newman,
  Conselice, Koekemoer, Lotz, \& Mozena}]{Freeman2013}
Freeman P.~E., Izbicki R., Lee A.~B., Newman J.~A., Conselice C.~J., Koekemoer
  A.~M., Lotz J.~M., Mozena M., 2013, \mnras, 434, 282

\bibitem[{Galametz {et~al}\mbox{.}(2013)Galametz, Grazian, Fontana, Ferguson,
  Ashby, Barro, Castellano, Dahlen, Donley, Faber, Grogin, Guo, Huang,
  Kocevski, Koekemoer, Lee, McGrath, Peth, Willner, Almaini, Cooper, Cooray,
  Conselice, Dickinson, Dunlop, Fazio, Foucaud, Gardner, Giavalisco, Hathi,
  Hartley, Koo, Lai, de~Mello, McLure, Lucas, Paris, Pentericci, Santini,
  Simpson, Sommariva, Targett, Weiner, \& Wuyts}]{Galametz2013}
Galametz A. {et~al.}, 2013, \apjs, 206, 10

\bibitem[{Glasser(1962)}]{Glasser1962}
Glasser G.~J., 1962, J. Am. Stat. Assoc., 57, 648

\bibitem[{Gordon {et~al}\mbox{.}(2003)Gordon, Clayton, Misselt, Landolt, \&
  Wolff}]{Gordon2003}
Gordon K.~D., Clayton G.~C., Misselt K.~A., Landolt A.~U., Wolff M.~J., 2003,
  \apj, 594, 279

\bibitem[{Governato {et~al}\mbox{.}(2009)Governato, Brook, Brooks, Mayer,
  Willman, Jonsson, Stilp, Pope, Christensen, Wadsley, \&
  Quinn}]{Governato2009}
Governato F. {et~al.}, 2009, \mnras, 398, 312

\bibitem[{Governato {et~al}\mbox{.}(2004)Governato, Mayer, Wadsley, Gardner,
  Willman, Hayashi, Quinn, Stadel, \& Lake}]{Governato2004}
Governato F. {et~al.}, 2004, \apj, 607, 688

\bibitem[{Grogin {et~al}\mbox{.}(2011)Grogin, Kocevski, Faber, Ferguson,
  Koekemoer, Riess, Acquaviva, Alexander, Almaini, \& Ashby}]{Grogin2011}
Grogin N.~A. {et~al.}, 2011, \apjs, 197, 35

\bibitem[{Groves {et~al}\mbox{.}(2008)Groves, Dopita, Sutherland, Kewley,
  Fischera, Leitherer, Brandl, \& van Breugel}]{groves08}
Groves B., Dopita M., Sutherland R., Kewley L., Fischera J., Leitherer C.,
  Brandl B., van Breugel W., 2008, \apjs, 176, 438

\bibitem[{Guedes {et~al}\mbox{.}(2011)Guedes, Callegari, Madau, \&
  Mayer}]{Guedes2011}
Guedes J., Callegari S., Madau P., Mayer L., 2011, \apj, 742, 76

\bibitem[{Guo {et~al}\mbox{.}(2015)Guo, Ferguson, Bell, Koo, Conselice,
  Giavalisco, Kassin, Lu, Lucas, Mandelker, McIntosh, Primack, Ravindranath,
  Barro, Ceverino, Dekel, Faber, Fang, Koekemoer, Noeske, Rafelski, \&
  Straughn}]{Guo2015}
Guo Y. {et~al.}, 2015, \apj, 800, 39

\bibitem[{Guo {et~al}\mbox{.}(2012)Guo, Giavalisco, Ferguson, Cassata, \&
  Koekemoer}]{Guo2012b}
Guo Y., Giavalisco M., Ferguson H.~C., Cassata P., Koekemoer A.~M., 2012, \apj,
  757, 120

\bibitem[{Hayward {et~al}\mbox{.}(2013)Hayward, Narayanan, Keres, Jonsson,
  Hopkins, Cox, \& Hernquist}]{Hayward2012}
Hayward C.~C., Narayanan D., Keres D., Jonsson P., Hopkins P.~F., Cox T.~J.,
  Hernquist L., 2013, \mnras, 428, 2529

\bibitem[{Hopkins {et~al}\mbox{.}(2013)Hopkins, Keres, Onorbe, Faucher-Giguere,
  Quataert, Murray, \& Bullock}]{Hopkins2013a}
Hopkins P.~F., Keres D., Onorbe J., Faucher-Giguere C.-A., Quataert E., Murray
  N., Bullock J.~S., 2013, eprint arXiv:1311.2073

\bibitem[{Hunter(2007)}]{Hunter:2007}
Hunter J.~D., 2007, Comput. Sci. \& Eng., 9, 90

\bibitem[{James {et~al}\mbox{.}(2002)James, Dunne, Eales, \&
  Edmunds}]{James:2002}
James A., Dunne L., Eales S., Edmunds M.~G., 2002, \mnras, 335, 753

\bibitem[{Johnson \& Morgan(1953)}]{johnsonmorgan53}
Johnson H., Morgan W., 1953, \apj, 117, 313

\bibitem[{Jonsson(2006)}]{jonsson06}
Jonsson P., 2006, \mnras, 372, 2

\bibitem[{Jonsson {et~al}\mbox{.}(2006)Jonsson, Cox, Primack, \&
  Somerville}]{Jonsson:2006}
Jonsson P., Cox T.~J., Primack J.~R., Somerville R.~S., 2006, \apj, 637, 255

\bibitem[{Jonsson, Groves \& Cox(2010)Jonsson, Groves, \& Cox}]{jonsson09}
Jonsson P., Groves B., Cox T., 2010, \mnras, 403, 17

\bibitem[{Jonsson \& Primack(2010)}]{Jonsson:2010gpu}
Jonsson P., Primack J.~R., 2010, New Astron., 15, 509

\bibitem[{Kartaltepe {et~al}\mbox{.}(2014)Kartaltepe, Mozena, Kocevski,
  McIntosh, Lotz, Bell, Faber, Ferguson, Koo, Bassett, Bernyk, Blancato,
  Bournaud, Cassata, Castellano, Cheung, Conselice, Croton, Dahlen, de Mello,
  DeGroot, Donley, Guedes, Grogin, Hathi, Hilton, Hollon, Inami, Kassin,
  Koekemoer, Lani, Liu, Lucas, Martig, McGrath, McPartland, Mobasher, Morlock,
  Mutch, O'Leary, Peth, Pforr, Pillepich, Poole, Rizer, Rosario, Soto,
  Straughn, Telford, Sunnquist, Weiner, \& Wuyts}]{Kartaltepe2014}
Kartaltepe Â. {et~al.}, 2014, eprint arXiv:1401.2455

\bibitem[{Kassin {et~al}\mbox{.}(2012)Kassin, Weiner, Faber, Gardner, Willmer,
  Coil, Cooper, Devriendt, Dutton, Guhathakurta, Koo, Metevier, Noeske, \&
  Primack}]{Kassin2012}
Kassin S.~A. {et~al.}, 2012, \apj, 758, 106

\bibitem[{Kauffmann {et~al}\mbox{.}(2003)Kauffmann, Heckman, White, Charlot,
  Tremonti, Peng, Seibert, Brinkmann, Nichol, SubbaRao, \&
  York}]{Kauffmann2003}
Kauffmann G. {et~al.}, 2003, \mnras, 341, 54

\bibitem[{Kennicutt(1998)}]{Kennicutt:1998}
Kennicutt R.~C., 1998, \apj, 498, 541

\bibitem[{Kim {et~al}\mbox{.}(2014)Kim, Abel, Agertz, Bryan, Ceverino,
  Christensen, Conroy, Dekel, Gnedin, Goldbaum, Guedes, Hahn, Hobbs, Hopkins,
  Hummels, Iannuzzi, Keres, Klypin, Kravtsov, Krumholz, Kuhlen, Leitner, Madau,
  Mayer, Moody, Nagamine, Norman, Onorbe, O'Shea, Pillepich, Primack, Quinn,
  Read, Robertson, Rocha, Rudd, Shen, Smith, Szalay, Teyssier, Thompson,
  Todoroki, Turk, Wadsley, Wise, \& Zolotov}]{Kim2014}
Kim J.-h. {et~al.}, 2014, \apjs, 210, 14

\bibitem[{Koekemoer {et~al}\mbox{.}(2011)Koekemoer, Faber, Ferguson, Grogin,
  Kocevski, Koo, Lai, Lotz, Lucas, \& McGrath}]{Koekemoer2011}
Koekemoer A.~M. {et~al.}, 2011, \apjs, 197, 36

\bibitem[{Komatsu {et~al}\mbox{.}(2009)Komatsu, Dunkley, Nolta, Bennett, Gold,
  Hinshaw, Jarosik, Larson, Limon, Page, Spergel, Halpern, Hill, Kogut, Meyer,
  Tucker, Weiland, Wollack, \& Wright}]{Komatsu2009}
Komatsu E. {et~al.}, 2009, \apjs, 180, 330

\bibitem[{Kravtsov(2003)}]{Kravtsov2003}
Kravtsov A.~V., 2003, \apj, 590, L1

\bibitem[{Kravtsov, Klypin \& Khokhlov(1997)Kravtsov, Klypin, \&
  Khokhlov}]{Kravtsov1997}
Kravtsov A.~V., Klypin A.~A., Khokhlov A.~M., 1997, \apjs, 111, 73

\bibitem[{Krist, Hook \& Stoehr(2011)Krist, Hook, \& Stoehr}]{Krist2011}
Krist J.~E., Hook R.~N., Stoehr F., 2011, in Opt. Model. Perform. Predict. V.
  Ed. by Kahan, Kahan M.~A., ed., Vol. 8127, pp. 81270J--81270J--16

\bibitem[{Kroupa(2001)}]{Kroupa:2001}
Kroupa P., 2001, \mnras, 322, 231

\bibitem[{Lanz {et~al}\mbox{.}(2014)Lanz, Hayward, Zezas, Smith, Ashby,
  Brassington, Fazio, \& Hernquist}]{Lanz2014}
Lanz L., Hayward C.~C., Zezas A., Smith H.~A., Ashby M. L.~N., Brassington N.,
  Fazio G.~G., Hernquist L., 2014, \apj, 785, 39

\bibitem[{Lee {et~al}\mbox{.}(2013)Lee, Giavalisco, Williams, Guo, Lotz, {Van
  der Wel}, Ferguson, Faber, Koekemoer, Grogin, Kocevski, Conselice, Wuyts,
  Dekel, Kartaltepe, \& Bell}]{Lee2013}
Lee B. {et~al.}, 2013, \apj, 774, 47

\bibitem[{Leitherer {et~al}\mbox{.}(1999)Leitherer, Schaerer, Goldader,
  Delgado, Robert, Kune, de~Mello, Devost, \& Heckman}]{leitherer99}
Leitherer C. {et~al.}, 1999, \apjs, 123, 3

\bibitem[{Lotz {et~al}\mbox{.}(2008{\natexlab{a}})Lotz, Davis, Faber,
  Guhathakurta, Gwyn, Huang, Koo, {Le Floc'h}, Lin, Newman, Noeske, Papovich,
  Willmer, Coil, Conselice, Cooper, Hopkins, Metevier, Primack, Rieke, \&
  Weiner}]{lotz08_hst}
Lotz J. {et~al.}, 2008{\natexlab{a}}, \apj, 672, 177

\bibitem[{Lotz {et~al}\mbox{.}(2008{\natexlab{b}})Lotz, Jonsson, Cox, \&
  Primack}]{lotz08}
Lotz J., Jonsson P., Cox T., Primack J., 2008{\natexlab{b}}, \mnras, 391, 1137

\bibitem[{Lotz {et~al}\mbox{.}(2010{\natexlab{a}})Lotz, Jonsson, Cox, \&
  Primack}]{lotz10}
Lotz J., Jonsson P., Cox T., Primack J., 2010{\natexlab{a}}, \mnras, 404, 575

\bibitem[{Lotz {et~al}\mbox{.}(2011)Lotz, Jonsson, Cox, Croton, Primack,
  Somerville, \& Stewart}]{Lotz2011}
Lotz J.~M., Jonsson P., Cox T.~J., Croton D., Primack J.~R., Somerville R.~S.,
  Stewart K., 2011, \apj, 742, 103

\bibitem[{Lotz {et~al}\mbox{.}(2010{\natexlab{b}})Lotz, Jonsson, Cox, \&
  Primack}]{Lotz2010}
Lotz J.~M., Jonsson P., Cox T.~J., Primack J.~R., 2010{\natexlab{b}}, \mnras,
  404, 590

\bibitem[{Lotz {et~al}\mbox{.}(2006)Lotz, Madau, Giavalisco, Primack, \&
  Ferguson}]{Lotz2006}
Lotz J.~M., Madau P., Giavalisco M., Primack J., Ferguson H.~C., 2006, \apj,
  636, 592

\bibitem[{Lotz, Primack \& Madau(2004)Lotz, Primack, \& Madau}]{Lotz2004}
Lotz J.~M., Primack J., Madau P., 2004, \aj, 128, 163

\bibitem[{Lupton {et~al}\mbox{.}(2004)Lupton, Blanton, Fekete, Hogg,
  O'Mullane, Szalay, \& Wherry}]{Lupton2004}
Lupton R., Blanton M.~R., Fekete G., Hogg D.~W., O'Mullane W., Szalay A.,
  Wherry N., 2004, \pasp, 116, 133

\bibitem[{Mandelbaum {et~al}\mbox{.}(2006)Mandelbaum, Seljak, Kauffmann,
  Hirata, \& Brinkmann}]{Mandelbaum2006}
Mandelbaum R., Seljak U., Kauffmann G., Hirata C.~M., Brinkmann J., 2006,
  \mnras, 368, 715

\bibitem[{Mandelker {et~al}\mbox{.}(2014)Mandelker, Dekel, Ceverino, Tweed,
  Moody, \& Primack}]{Mandelker2014}
Mandelker N., Dekel A., Ceverino D., Tweed D., Moody C.~E., Primack J., 2014,
  \mnras, 443, 3675

\bibitem[{Marinacci, Pakmor \& Springel(2013)Marinacci, Pakmor, \&
  Springel}]{Marinacci2013}
Marinacci F., Pakmor R., Springel V., 2013, \mnras, 437, 1750

\bibitem[{Moody {et~al}\mbox{.}(2014)Moody, Guo, Mandelker, Ceverino, Mozena,
  Koo, Dekel, \& Primack}]{Moody2014}
Moody C.~E., Guo Y., Mandelker N., Ceverino D., Mozena M., Koo D.~C., Dekel A.,
  Primack J., 2014, \mnras, 444, 1389

\bibitem[{Moster {et~al}\mbox{.}(2010)Moster, Somerville, Maulbetsch, van~den
  Bosch, Macci\`{o}, Naab, \& Oser}]{Moster2010}
Moster B.~P., Somerville R.~S., Maulbetsch C., van~den Bosch F.~C., Macci\`{o}
  A.~V., Naab T., Oser L., 2010, \apj, 710, 903

\bibitem[{Narayanan {et~al}\mbox{.}(2010)Narayanan, Hayward, Cox, Hernquist,
  Jonsson, Younger, \& Groves}]{narayanan10_smg}
Narayanan D., Hayward C., Cox T., Hernquist L., Jonsson P., Younger J., Groves
  B., 2010, \mnras, 401, 1613

\bibitem[{Omand, Balogh \& Poggianti(2014)Omand, Balogh, \&
  Poggianti}]{Omand2014}
Omand C. M.~B., Balogh M.~L., Poggianti B.~M., 2014, \mnras, 440, 843

\bibitem[{Patel {et~al}\mbox{.}(2012)Patel, Holden, Kelson, Franx, van~der Wel,
  \& Illingworth}]{Patel2012}
Patel S.~G., Holden B.~P., Kelson D.~D., Franx M., van~der Wel A., Illingworth
  G.~D., 2012, \apj, 748, L27

\bibitem[{Pedrosa, Tissera \& {De Rossi}(2014)Pedrosa, Tissera, \& {De
  Rossi}}]{Pedrosa2014}
Pedrosa S.~E., Tissera P.~B., {De Rossi} M.~E., 2014, \aap, 567, A47

\bibitem[{Porter {et~al}\mbox{.}(2014)Porter, Somerville, Primack, \&
  Johansson}]{Porter2014}
Porter L.~A., Somerville R.~S., Primack J.~R., Johansson P.~H., 2014, \mnras,
  444, 942

\bibitem[{Robertson {et~al}\mbox{.}(2006)Robertson, Bullock, Cox, {Di Matteo},
  Hernquist, Springel, \& Yoshida}]{robertson06}
Robertson B., Bullock J., Cox T., {Di Matteo} T., Hernquist L., Springel V.,
  Yoshida N., 2006, \apj, 645, 986

\bibitem[{Rocha {et~al}\mbox{.}(2007)Rocha, Jonsson, Primack, \&
  Cox}]{Rocha2007}
Rocha M., Jonsson P., Primack J.~R., Cox T.~J., 2007, \mnras, 383, 1281

\bibitem[{Scannapieco {et~al}\mbox{.}(2010)Scannapieco, Gadotti, Jonsson, \&
  White}]{scannapieco10}
Scannapieco C., Gadotti D., Jonsson P., White S., 2010, \mnras, L99

\bibitem[{Scannapieco {et~al}\mbox{.}(2012)Scannapieco, Wadepuhl, Parry,
  Navarro, Jenkins, Springel, Teyssier, Carlson, Couchman, Crain, Vecchia,
  Frenk, Kobayashi, Monaco, Murante, Okamoto, Quinn, Schaye, Stinson, Theuns,
  Wadsley, White, \& Woods}]{Scannapieco2012}
Scannapieco C. {et~al.}, 2012, \mnras, 423, 1726

\bibitem[{Snyder {et~al}\mbox{.}(2011)Snyder, Cox, Hayward, Hernquist, \&
  Jonsson}]{snyder11a}
Snyder G., Cox T., Hayward C., Hernquist L., Jonsson P., 2011, \apj, 741, 77

\bibitem[{Snyder {et~al}\mbox{.}(2013)Snyder, Hayward, Sajina, Jonsson, Cox,
  Hernquist, Hopkins, \& Yan}]{Snyder2013}
Snyder G.~F., Hayward C.~C., Sajina A., Jonsson P., Cox T.~J., Hernquist L.,
  Hopkins P.~F., Yan L., 2013, \apj, 768, 168

\bibitem[{Stinson {et~al}\mbox{.}(2012)Stinson, Brook, Maccio, Wadsley, Quinn,
  \& Couchman}]{Stinson2012}
Stinson G.~S., Brook C., Maccio A.~V., Wadsley J., Quinn T.~R., Couchman H.
  M.~P., 2012, \mnras, 428, 129

\bibitem[{Torrey {et~al}\mbox{.}(2015)Torrey, Snyder, Vogelsberger, Hayward,
  Genel, Sijacki, Springel, Hernquist, Nelson, Kriek, Pillepich, Sales, \&
  McBride}]{Torrey2015}
Torrey P. {et~al.}, 2015, \mnras, 447, 2753

\bibitem[{Torrey {et~al}\mbox{.}(2014)Torrey, Vogelsberger, Genel, Sijacki,
  Springel, \& Hernquist}]{Torrey2014}
Torrey P., Vogelsberger M., Genel S., Sijacki D., Springel V., Hernquist L.,
  2014, \mnras, 438, 1985

\bibitem[{Trujillo-Gomez {et~al}\mbox{.}(2013)Trujillo-Gomez, Klypin, Colin,
  Ceverino, Arraki, \& Primack}]{Trujillo-Gomez2013}
Trujillo-Gomez S., Klypin A., Colin P., Ceverino D., Arraki K., Primack J.,
  2013, eprint arXiv:1311.2910

\bibitem[{Turk {et~al}\mbox{.}(2011)Turk, Smith, Oishi, Skory, Skillman, Abel,
  \& Norman}]{Turk2011}
Turk M.~J., Smith B.~D., Oishi J.~S., Skory S., Skillman S.~W., Abel T., Norman
  M.~L., 2011, \apjs, 192, 9

\bibitem[{van~der Wel {et~al}\mbox{.}(2014)van~der Wel, Chang, Bell, Holden,
  Ferguson, Giavalisco, Rix, Skelton, Whitaker, Momcheva, Brammer, Kassin,
  Martig, Dekel, Ceverino, Koo, Mozena, van Dokkum, Franx, Faber, \&
  Primack}]{VanderWel2014}
van~der Wel A. {et~al.}, 2014, \apjl, 792, L6

\bibitem[{Weingartner \& Draine(2001)}]{wd01}
Weingartner J., Draine B., 2001, \apj, 548, 296

\bibitem[{Wellons {et~al}\mbox{.}(2014)Wellons, Torrey, Ma, Rodriguez-Gomez,
  Vogelsberger, Kriek, van Dokkum, Nelson, Genel, Pillepich, Springel, Sijacki,
  Snyder, Nelson, Sales, \& Hernquist}]{Wellons2014}
Wellons S. {et~al.}, 2014, eprint arXiv:1411.0667, MNRAS accepted

\bibitem[{Whitaker {et~al}\mbox{.}(2014)Whitaker, Franx, Leja, van Dokkum,
  Henry, Skelton, Fumagalli, Momcheva, Brammer, Labbe, Nelson, \&
  Rigby}]{Whitaker2014}
Whitaker K.~E. {et~al.}, 2014, eprint arXiv:1407.1843

\bibitem[{Whitaker {et~al}\mbox{.}(2013)Whitaker, van Dokkum, Brammer,
  Momcheva, Skelton, Franx, Kriek, Labb\'{e}, Fumagalli, Lundgren, Nelson,
  Patel, \& Rix}]{Whitaker2013}
Whitaker K.~E. {et~al.}, 2013, \apjl, 770, L39

\bibitem[{Williams {et~al}\mbox{.}(2009)Williams, Quadri, Franx, van Dokkum, \&
  Labb\'{e}}]{Williams2009}
Williams R.~J., Quadri R.~F., Franx M., van Dokkum P., Labb\'{e} I., 2009,
  \apj, 691, 1879

\bibitem[{Wuyts {et~al}\mbox{.}(2011)Wuyts, {F\"{o}rster Schreiber}, van~der
  Wel, Magnelli, Guo, Genzel, Lutz, Aussel, Barro, Berta, Cava,
  Graci\'{a}-Carpio, Hathi, Huang, Kocevski, Koekemoer, Lee, {Le Floc'h},
  McGrath, Nordon, Popesso, Pozzi, Riguccini, Rodighiero, Saintonge, \&
  Tacconi}]{Wuyts2011}
Wuyts S. {et~al.}, 2011, \apj, 742, 96

\bibitem[{Zolotov {et~al}\mbox{.}(2014)Zolotov, Dekel, Mandelker, Tweed, Inoue,
  DeGraf, Ceverino, \& Primack}]{Zolotov2014}
Zolotov A., Dekel A., Mandelker N., Tweed D., Inoue S., DeGraf C., Ceverino D.,
  Primack J., 2014, eprint arXiv:1412.4783

\end{thebibliography}
\bsp


\end{document}